 \documentclass[final, 3p, 11pt, times]{elsarticle}

\usepackage{amsmath}
\usepackage{subfig}

\usepackage{amssymb}

\journal{Journal of Mechanics and Physics of Solids}

\newcommand{\bff}{{\bf f}}

\newcommand{\bfu}{{\bf u}}

\newcommand{\bfC}{{\bf C}}
\newcommand{\bfE}{{\bf E}}
\newcommand{\bfF}{{\bf F}}

\newcommand{\bfI}{{\bf I}}
\newcommand{\bfK}{{\bf K}}
\newcommand{\bfM}{{\bf M}}

\newcommand{\bfS}{{\bf S}}

\newcommand{\beq}{\begin{equation}}
\newcommand{\eeq}{\end{equation}}
\newcommand{\beqs}{\begin{eqnarray}}
\newcommand{\eeqs}{\end{eqnarray}}
\newcommand{\beql}{\begin{equation} \label}
\newcommand{\half}{\frac{1}{2}}

\begin{document}

\begin{frontmatter}



\title{Nonlinear Waves in Lattice Materials: Adaptively Augmented Directivity and Functionality Enhancement by Modal Mixing}


\author{R. Ganesh}
\ead{ramak015@umn.edu}

\author{S. Gonella \corref{lad}}
\ead{sgonella@umn.edu}
\cortext[lad]{Corresponding Author \\ \indent\indent Address: 500 Pillsbury Drive S.E., Minneapolis, MN, 55455-0116; Voice: 612-625-0866; Fax: 612-626-7750}

\address{Department of Civil, Environmental, and Geo- Engineering, University of Minnesota, Minneapolis, MN, United States}

\begin{abstract}
The motive of this work is to understand the complex spatial characteristics of finite-amplitude elastic wave propagation in periodic structures and leverage the unique opportunities offered by nonlinearity to activate complementary functionalities and design adaptive spatial wave manipulators. The underlying assumption is that the magnitude of wave propagation is small with respect to the length scale of the structure under consideration, albeit large enough to elicit the effects of finite-deformation. We demonstrate that the interplay of dispersion, nonlinearity and modal complexity involved in the generation and propagation of higher harmonics gives rise to secondary wave packets that feature multiple characteristics, one of which conforms to the dispersion relation of the corresponding linear structure. This provides an opportunity to engineer desired wave characteristics through a geometric and topological design of the unit cell, and results in the ability to activate complementary functionalities, typical of high frequency regimes, while operating at low frequencies of excitation - an effect seldom observed in linear periodic structures. The ability to design adaptive switches is demonstrated here using lattice configurations whose response is characterized by geometric and/or material nonlinearities. 
\end{abstract}


\end{frontmatter}


\section{Introduction}
Periodic structures are defined as media featuring a repetitive pattern of discrete elements referred to as unit cells. The repetitive pattern can be established through a periodic modulation of the geometric or material properties, or can manifest as a result of certain boundary conditions imposed on the structure \cite{Hussein2014}. Periodic structures are known for their ability to inhibit the propagation of waves within certain frequency bands (referred to as bandgaps), thereby functioning as mechanical filters or waveguides \cite{Kushwaha1993,Sigalas1995,Martinez-Sala1995,Hladky-Hennion2007}. In addition, the velocity of wave propagation is also dependent on the frequency and direction of propagation, which leads to significant frequency-dependent energy directivity \cite{Langley1996,Phani2006,Spadoni2009}.\newline
\indent Despite these unique characteristics, the practical applicability of \textit{linear} periodic structures is limited by their inherent passivity; i.e., for a given geometry, material and/or boundary conditions, the bandgaps and spatial directivity patterns are fixed. In this regard, finite-deformation elastic effects have been explored as a possible avenue to achieve tunability of the spectro-spatial characteristics of periodic structures. For example, nonlinearity has been shown to impart amplitude-dependent characteristics to the dispersion relation, resulting in the modification of the propagation and attenuation zones (bandgaps) of the structure. For weakly nonlinear systems, where the contribution of the nonlinear terms to the response is much smaller than that of the linear terms, the corrected dispersion relation can be determined through the application of the $Lindstedt-Poincar\acute{e}$ perturbation technique \cite{Askar1973,Chakraborty2001,Narisetti2010}. In addition, the spatial characteristics of these systems, described by the phase and group velocity contours, have also been shown to feature amplitude-dependent behavior \cite{Manktelow2013}. While the bulk of the literature is focused on the weakly nonlinear regime, some studies have also considered strongly nonlinear systems. For example, \citet{Abedinnasab2013} derived exact dispersion relations for nonlinear wave propagation in rods and beams, while \citet{Narisetti2012} obtained dispersion relations for strongly nonlinear periodic structures using the Harmonic Balance method.\newline
\indent Tunability of small-amplitude wave characteristics has also been studied in a number of conventional and unconventional media capable of undergoing hyperelastic deformations. In isotropic solids, this effect is typically captured in the form of a stress dependence of the wave velocity, which was used by \citet{Hughes1953} to determine experimentally the second-order elastic constants of materials. Along these lines, \citet{Bertoldi2008} considered an elastomeric periodic structure featuring square arrays of circular holes, and studied the propagation characteristics of small-amplitude waves superimposed to large static precompressions. They observed that the instabilities of the system triggered under the application of large static forces produced pattern transformations, thus resulting in a modification of the unit cell and in the appearance of new bandgaps in the dispersion diagram. This type of reversible pattern transformation has been exploited to design structures with tunable bandgaps and directional features \cite{Wang2014, Shan2014}. Similarly, confined precompressed monoatomic granular chains, which can be modeled as FPU lattices in the limit of strong precompression (for dynamic displacements that are much smaller than the static precompression applied to the structure), feature force-dependent dispersion relations. Therefore, by controlling the precompression force, the cutoff frequency of the chains can be tuned, as demonstrated experimentally for a $1$D diatomic granular crystal by \citet{Boechler2011a}.\newline
\indent Another well-known effect of nonlinearity is the generation of super- and sub-harmonics in the response, due to the self-interaction of the applied excitation. In dispersive systems, the ability to propagate higher harmonics is governed by the availability of dispersion branches in the frequency range of the activated harmonics, which was demonstrated for nonlinear Lamb wave propagation in isotropic plates by \citet{Deng1999}  and \citet{DeLima2003}. In general, the nonlinearly generated higher harmonics display oscillatory, bounded characteristics corresponding to the dispersion relation of the structure under consideration. However, these oscillations become unbounded when resonance conditions are satisfied, i.e., when the fundamental and the higher harmonics have the same phase velocity. This cumulative generation of harmonics has been observed experimentally in isotropic elastic plates \cite{Deng2005,Matlack2011}, thereby providing a route to experimentally measure nonlinear parameters using guided wave testing \cite{Deng2007,Jacobs2007}. For acoustic waves, \citet{Bradley1995} studied the generation of higher harmonics in a nonlinear periodic waveguide by modeling the nonlinear fluid using the modified Westervelt equation, and demonstrated experimentally the generation of spatially beating (bounded oscillation) second harmonic in an air-filled waveguide. The use of the Westervelt equation as a means to understand the interaction between dispersion and nonlinearity in fluid systems has been extensively studied by various authors, and an excellent review is  available in \cite{hamilton1998}.  For elastic wave propagation in strongly precompressed monoatomic granular chains, \citet{Morcillo2013} obtained an expression for the amplitude of the second harmonic generated due to nonlinearity. The generation of higher harmonics in a diatomic chain has also been studied theoretically and experimentally by \citet{Cabaret2012}. In all these cases, the generated second harmonic was considered to be non-resonant with the fundamental harmonic. The resonant case needs to be treated separately using the condition of three-wave-interaction, which has been carried out for generic dispersive hyperbolic differential equations \cite{Schneider2005} and periodic lattices \cite{Konotop1996b,Goncalves2000}. In addition, nonlinearity also gives rise to sub-harmonics, which have been studied in precompressed monoatomic granular chains by \citet{Tournat2004}.\newline
\indent The presence of spatial periodicity in lattices gives rise to additional modal complexity, whose effect on nonlinear harmonic generation has seldom been studied in a systematic fashion. We have previously explored this type of phenomena in the context of granular phononic crystals \cite{GG15}, where we have shown that the generation of higher harmonics induces jumps in the response across propagation modes, and results in a response that features a mixture of modes and simultaneous activation of complementary functionalities. In this work, we expand the framework of \cite{GG15} by providing a systematic mathematical description of the problem, and by taking the concept of modal mixing to the terrain of (cellular) continua - a realm with additional modeling complexity as well as richer opportunities for wave manipulation. The mathematical framework is obtained by employing a multiple scales expansion that captures the spatiotemporal characteristics of compact traveling wave packets \cite{Taniuti1969}. While this expansion has been utilized to explore a wide variety of systems \cite{DeSterke1988,Huang1993,Huang1998,Chirilus2012,Chong2013}, its complete implications in the context of elastic lattice materials have not been explored. In the first part of the work, we apply the multiple scales expansion to a simple diatomic nonlinear spring-mass chain, which is the simplest tractable mechanical system that incorporates the three interplaying characteristics of periodic solids that we intend to exploit for metamaterial design: nonlinearity, dispersion, and modal complexity. Here, we show that the spatial characteristics of the higher harmonics can be modified by manipulating the properties of the linearized unit cell, thus resulting in a modally augmented wave propagation scenario (with respect to that of a linear chain). Then, we shift our target to complex cellular solids featuring geometric and/or material nonlinearities and we demonstrate that the spatial characteristics of the higher harmonics can be engineered to enrich the directivity landscape of these structures and activate functionalities complementary to that of the fundamental harmonic, ultimately resulting in opportunities for adaptive spatial manipulation.
\section{The illustrative case of diatomic spring-mass chains}
\label{sec2}
In this section, we present the mathematical framework necessary to study the propagation of finite-amplitude elastic waves in periodic structures in a way that fully captures the manifestation of nonlinearity at all spatial and temporal scales. The formulation is given using the illustrative example of a diatomic chain which, due to its simplicity, allows for a tractable analytical derivation and, due to its abstractness, provides a general framework that can be extended, without loss of generality, to more complex structural systems (as done in the following sections). The multiple scales expansion depends on a non-trivial representation for traveling wave packets, which is presented here using the example of a generic $1$D linear dispersive system.
\subsection{Multiple scales representation of a traveling wave packet}
\begin{figure}[!htb]
\centering
\includegraphics[scale = 0.12]{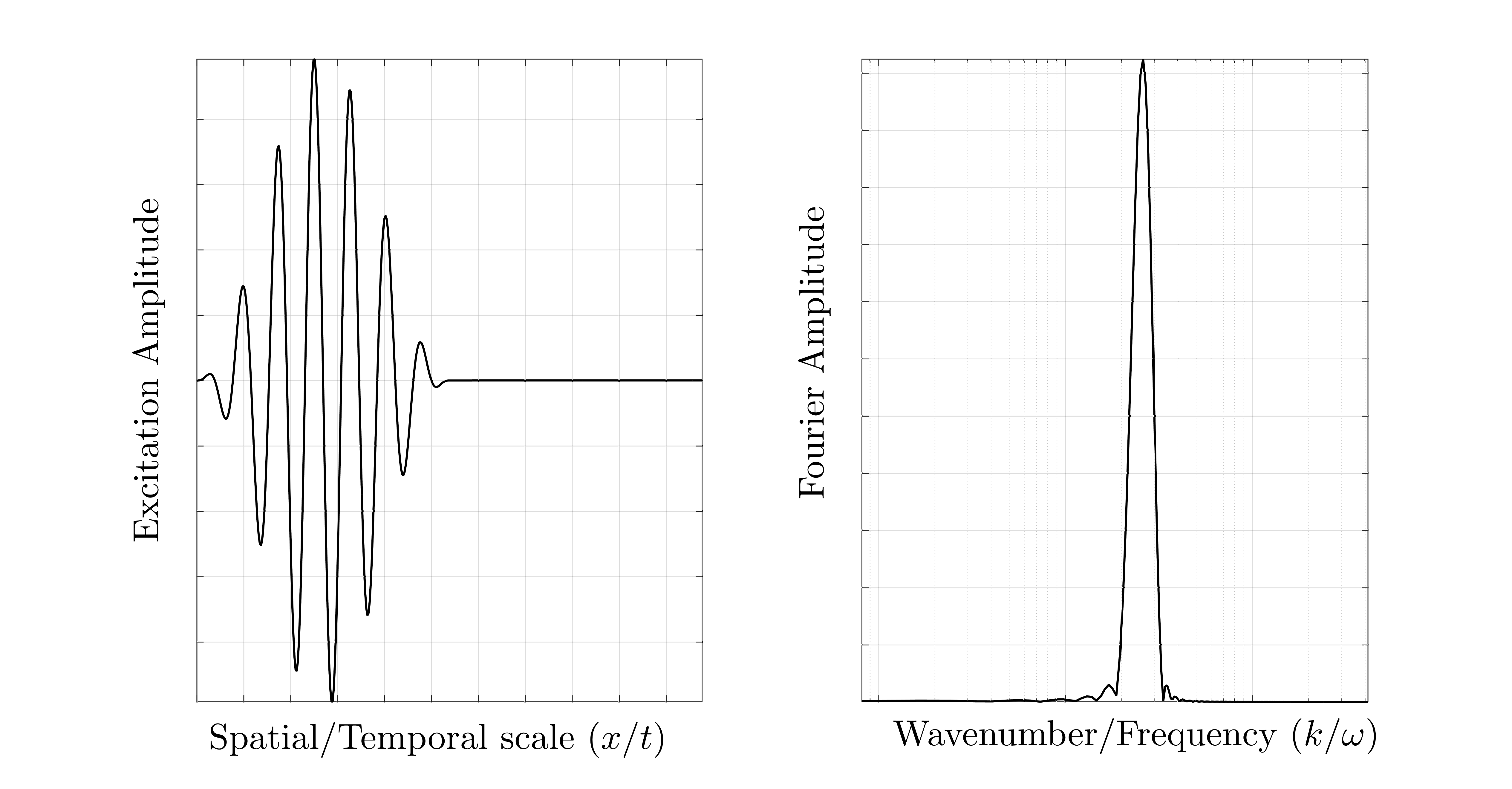}
\caption{Spatial and spectral representation of a tone burst.} 
\label{fig2.1}
\end{figure}

The Fourier representation of a $1$D traveling wave in a dispersive, linear system is given by 
\begin{equation}
\label{eq2.0}
u(x,t) = \int_{-\infty}^{\infty} F(k)e^{i\left(kx-\omega(k) t\right)} dk ,
\end{equation}
where $x$ and $t$ are the spatial and temporal variables, while $k$ (wavenumber) and $\omega$ (frequency) are the corresponding spectral variables. $F(k)$ is the Fourier amplitude corresponding to a given wavenumber $k$, and $\omega(k)$ is determined from the dispersion relation for the system under consideration. The spectral signature of a narrow-band wave packet, as shown in fig.~\ref{fig2.1}, consists of a carrier frequency  and wavenumber $(\omega_0, k_0)$ modulated by a long-wavelength envelope, such that the total spectral signature is limited to a small band of frequencies ($\Delta \omega, \Delta k$). For such conditions, eqn.~(\ref{eq2.0}) can be simplified as 
\begin{equation}
\label{eq2.1}
u(x,t) = \int_{k_0-\Delta k_0}^{k_0+\Delta k_0} F(k)e^{i\left(kx-\omega(k) t\right)} dk\,,
\end{equation}
where $F(k)$ is assumed to be non-zero only within a limited band of wavenumbers $[k_0-\Delta k,k_0+\Delta k]$. Assuming that the frequency band is small compared to the carrier frequency, the expression for $\omega$ can be simplified by employing a Taylor series expansion in the neighborhood of the carrier frequency $\omega_0(=\omega(k_0))$ as
\begin{equation}
\label{eq2.2}
\omega(k) = \omega(k_0+\chi) = \omega_0 + \frac{d\omega}{dk}\bigg|_{k_0}\chi + \half \frac{d^2\omega}{dk^2}\bigg|_{k_0}\chi^2 + \cdots.
\end{equation}
Substituting eqn.~(\ref{eq2.2}) in eqn.~(\ref{eq2.1}) and simplifying, we obtain
\begin{align}
u(x,t) &\approx e^{i\left(k_0x-\omega_0 t\right)} \int_{-\Delta k_0}^{\Delta k_0} F(k_0+\chi)e^{i\left(\chi x -\omega'(k_0)\chi t-\omega''(k_0) \chi^2 t\right)} d\chi \nonumber \\
 &= A\left(\chi(x-\omega'(k_0)t),\chi^2\omega''(k_0)t\right)\,\,e^{i\left(k_0x-\omega_0 t\right)}, \label{eq2.3}
\end{align}
where $A$ is a spatiotemporal function that also depends on the perturbation parameter $\chi$, and $\omega'= d\omega/dk$ and $\omega'' = d^2\omega/dk^2$ are determined from the dispersion relation. Since the expansion is carried out in the neighborhood of the carrier wavenumber, $\chi$ is a small parameter $\mathcal{O}(\varepsilon)$, where $\varepsilon<<1$. Therefore, $A$ can be considered as a function dependent only on slow scale variables $\mathcal{O}(\varepsilon)$ and higher. As a result, the traveling wave can be effectively represented using a multiple scales expansion, written in terms of three independent variables as 
\begin{equation}
u(x,t) = u(\theta,\xi,\tau) = A(\xi,\tau)\,e^{i\theta},
\label{eq2.32}
\end{equation}
where $\theta=k_0(x-c_pt)$ is the fast spatiotemporal variable, while $\xi=\varepsilon(x-c_gt)$ and $\tau=\varepsilon^2t$ are the slow scale variables. The fast oscillations $(e^{i(k_0x-\omega_0 t)})$ in eqn.~(\ref{eq2.32}) are invariant in the spatiotemporal space $x-c_pt$, where $c_p = \omega_0/k_0$ is the phase velocity of the propagating wave packet. This representation is consistent with the fundamental assumption that the harmonic characteristics of the burst can be assumed to be monochromatic.\newline
\indent The slow spatiotemporal scale $\xi$ corresponds to a frame of reference translating at a speed given by the group velocity $c_g (=d\omega/dk$), evaluated at the carrier frequency of the wave packet. This variable captures the dependence of the amplitude envelope of the wave packet on the group velocity. In addition, the envelope of the wave packet also depends on the slow temporal variable $\tau$, which implies that the envelope of the wave packet can be effectively described as a traveling wave that varies on a very slow time scale $\mathcal{O}(\varepsilon^2)$. This description of the wave packet is consistent with the well-known property of dispersive systems (for which $c_g\neq c_p$), where the velocity of the wave packet envelope is given by the group velocity, and whose distortion/broadening, which is observed in a slower temporal scale, is attributed to the variable speed of the different spectral components \cite{Remoissenet1999}.
\subsection{Multiple scales formulation for a diatomic chain}
\begin{figure}[!htb]
\centering
\includegraphics[scale = 0.25]{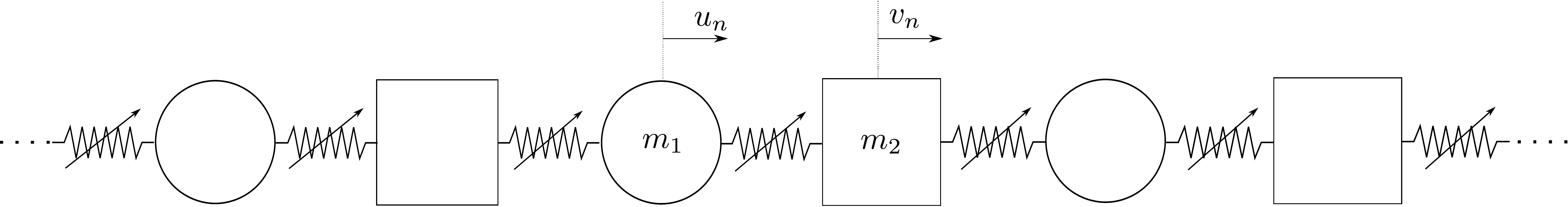}
\caption{Diatomic spring-mass chain}
\label{fig4.1}
\end{figure}

Consider a diatomic chain consisting of two types of masses $m_1$ and $m_2$ connected by identical springs, as shown in fig.~\ref{fig4.1}. The potential energy of the springs is assumed to be nonlinearly dependent on the change in length $(\delta)$ which, under the assumption of small displacements (with respect to the equilibrium position of the masses, $\delta\approx\mathcal{O}(\varepsilon)$), can be linearized using a Taylor series expansion. Thus, the restoring force $(f_r)$ in the springs can be expressed as
\begin{equation}
f_r(\delta)\,=\,K_2\,\delta\,+K_3\,\delta^{2} + K_4\,\delta^{3} +\mathcal{O}(\varepsilon^4),
\label{eqfc}
\end{equation}
where $K_2,\,K_3,\,K_4$ are the equivalent linear, quadratic, and cubic spring constants determined from the Taylor series expansion, and the number of terms that need to be retained in the expansion depend on the magnitude of the displacements. Using this constitutive relation, the governing equations for the motion of the $n^{th}$ set of masses ($m_1$ and $m_2$) can be derived as 
\begin{align}
\begin{bmatrix}
m_1 & 0 \\ 
0 & m_2
\end{bmatrix}
\begin{Bmatrix}
\ddot{u}_n \\ \ddot{v}_n
\end{Bmatrix} + 
\begin{bmatrix}
2K_2 & -K_2 \\ 
-K_2 & 2K_2
\end{bmatrix}
\begin{Bmatrix}
u_n \\ v_n
\end{Bmatrix}
&+ \begin{bmatrix}
0 & -K_2 \\ 
0 & 0
\end{bmatrix}
\begin{Bmatrix}
u_{n-1} \\ v_{n-1}
\end{Bmatrix} +
\begin{bmatrix}
0 & 0 \\ 
-K_2 & 0
\end{bmatrix}
\begin{Bmatrix}
u_{n+1} \\ v_{n+1}
\end{Bmatrix} \nonumber\\
+		
\begin{Bmatrix}
\varepsilon\,K_3\,[(u_{n}-v_{n-1})^{2}-(v_{n}-u_{n})^{2}] \\
\varepsilon\,K_3\,[(v_{n}-u_{n})^{2}-(u_{n+1}-v_{n})^{2}] 
\end{Bmatrix} 
&+ \begin{Bmatrix}
\varepsilon^2\,K_4\,[(u_{n}-v_{n-1})^{3}-(v_{n}-u_{n})^{3}] \\
\varepsilon^2\,K_4\,[(v_{n}-v_{n-1})^{3}-(u_{n+1}-v_{n})^{3}] 
\end{Bmatrix} =\,\begin{Bmatrix}
0 \\ 0
\end{Bmatrix},
\label{eq4.1} 
\end{align}
where $u_n,\, v_n$ are the displacements of the masses $m_1,\,m_2$ in the $n^{th}$ unit cell, respectively, and the approximation $u,v\approx\mathcal{O}(\varepsilon)$ has been incorporated to establish the relative orders of the terms in the equation. Since we are interested in delineating the characteristics of traveling wave packets, the solution to eqn.~(\ref{eq4.1}) can be written using a perturbative expansion, in the spirit of eqn.~(\ref{eq2.32}), as 
\begin{equation} 
\label{eq4.2}
\bfu_n = \begin{Bmatrix}
u_{n}(t) \\
v_{n}(t)
\end{Bmatrix} = \begin{Bmatrix}
\sum_{i=0}^{\infty} \varepsilon^i u^{i}(\theta_n,\xi_n,\tau)\\
\sum_{i=0}^{\infty} \varepsilon^i v^{i}(\theta_n,\xi_n,\tau)
\end{Bmatrix},
\end{equation}
where $\theta_n = kn-\omega(k) t$, $\xi_n = \varepsilon(n-c_gt)$, and $\tau = \varepsilon^2 t$ are the fast and slow scale variables introduced above, specialized here for a discrete system. Using eqn.~(\ref{eq4.2}), the acceleration of the masses can be expressed through a simple application of the chain rule in terms of partial derivatives with respect to the multiple scales, and the equations of motion can be separated by each order $(p)$ of the perturbation expansion as (details of the expansion are provided in \ref{appA})
\begin{equation}
\label{eq4}
\mathcal{O}(\varepsilon^p) :\;\;\; \omega^2 \bfM \frac{\partial^2 \bfu_{n}^p}{\partial \theta_n^2} + \sum_{i=n-1}^{n+1} \bfK_i \bfu_i^p = \bff^p,	
\end{equation}
where
\begin{equation*}
\bfM = \begin{bmatrix}
m_1 & 0 \\ 
0 & m_2
\end{bmatrix} ;\,\,
\bfK_n = \begin{bmatrix}
2K_2 & -K_2 \\ 
-K_2 & 2K_2
\end{bmatrix} ;\,\,
\bfK_{n-1} = \begin{bmatrix}
0 & -K_2 \\ 
0 & 0
\end{bmatrix} ;\,\,
\bfK_{n+1} = \begin{bmatrix}
0 & 0 \\ 
-K_2 & 0
\end{bmatrix}.
\end{equation*}
The forcing function $\bff^p$, at the first two orders of expansion, is
\begin{align}
\bff^0 &= \boldsymbol{0}, \nonumber \\
\bff^1 &= \begin{Bmatrix}
-2m_1\omega c_g \frac{\partial^2 u_{n}^{0}}{\partial \xi_n \partial \theta_n}-K_2\frac{\partial v_{n-1}^{0}}{\partial \xi_n}+K_3\,[(v_{n}^{0}-u_{n}^{0})^{2}-(u_{n}^{0}-v_{n-1}^{0})^{2}] \\
-2m_2\omega c_g \frac{\partial^2 v_{n}^{0}}{\partial \xi_n \partial \theta_n}+K_2\frac{\partial u_{n+1}^{0}}{\partial \xi_n}+K_3\,[(u_{n+1}^{0}-v_{n}^{0})^{2}-(v_{n}^{0}-u_{n}^{0})^{2}]
\end{Bmatrix} \label{eq4f1}.
\end{align}
\indent In essence, by introducing the multiple scales expansion, the governing equation for the weakly nonlinear spring-mass chain has been converted to a set of cascading linear differential equations, with $\mathcal{O}(\varepsilon)$ and higher being heterogeneous equations where the forcing function is determined from the solutions obtained at all the lower orders of expansion. Therefore, the solution at each order of expansion can be determined by solving the corresponding linear differential equation.\newline
\indent Considering eqn.~(\ref{eq4}) at $\mathcal{O}(1)$, the homogeneous solution can be expressed as 
\begin{equation}
\label{eqrefp0}
\bfu_n^0 = A_0(\xi_n,\tau) + A(\xi_n,\tau)\,\boldsymbol{\phi}\,e^{i\theta_n}+A^*(\xi_n,\tau)\,\boldsymbol{\phi^*}\,e^{-i\theta_n},	
\end{equation}
where $A_0$ and $A$ (referred to as slowly-varying envelopes) are constants that depend only on the slow scale variables $(\xi_n, \tau)$, and $(\cdot)^*$ refers to the complex conjugate of a variable. The functional dependence of the terms $A_0$ and $A$ on the slow scale variables $(\xi_n, \tau)$ is determined by solving the higher order equations at $\mathcal{O}(\varepsilon)$ and $\mathcal{O}(\varepsilon^2)$. While $A$ is the envelope modulation associated with the traveling wave packet discussed in eqn.~(\ref{eq2.32}), $A_0$ is a constant of integration whose existence, or lack thereof, depends on the nature of the problem.\newline
The relationship between $\omega$, $k$ and the modal vector $\boldsymbol \phi$ are obtained by enforcing Bloch conditions between neighboring cells on the fast scale variables, which results in the well-known wavenumber-dependent eigenvalue problem
\begin{equation}
\left(-\omega^2\begin{bmatrix}
m_1 & 0 \\ 
0 & m_2
\end{bmatrix}
+ \begin{bmatrix}
2K_2 & -K_2(1+e^{-ik}) \\ 
-K_2(1+e^{ik}) & 2K_2
\end{bmatrix}\right)\boldsymbol{\phi}=0,
\label{eq4.6}
\end{equation}
whose solution can be obtained analytically as 
\begin{align}
\omega_{ac/op} &= \sqrt{\frac{K_2}{m_1m_2}(m_1+m_2) \mp \frac{K_2}{m_1m_2} \sqrt{m_1^2 + m_2^2 + 2m_1m_2\cos{k}}}, \nonumber\\ 
\boldsymbol{\phi}_{ac/op} &= \begin{Bmatrix}
\frac{-(1+e^{-ik})m_2}{m1-m2\mp\sqrt{m_1^2 + m_2^2 + 2m_1m_2\cos{k}}}\\
1
\end{Bmatrix}.\nonumber 
\end{align}
For any wavenumber $k$, there exist two possible frequencies of wave propagation, one corresponding to the acoustic mode (with frequency $\omega_{ac}$ and modal vector $\boldsymbol{\phi}_{ac}$), and the other corresponding to the optical mode ($\omega_{op},\boldsymbol{\phi}_{op}$). In this work, the input excitation is assumed to fall in the range of the acoustic mode (simply referred to as $\omega, \phi$ henceforth).\newline
\indent At $\mathcal{O}(\varepsilon)$, the complete solution to eqn.~(\ref{eq4}) consists of a particular (forced) solution corresponding to the forcing function $\bff^1$, and the solution of the associated homogeneous problem. The forcing function is determined by substituting the zeroth order solution (determined in eqn.~(\ref{eqrefp0})) in eqn.~\ref{eq4f1}, which can be simplified as 
\begin{align}
\bff^1 = &\begin{Bmatrix}
-K_2 \frac{\partial A_{0}}{\partial \xi_n} \\ K_2 \frac{\partial A_{0}}{\partial \xi_n} 
\end{Bmatrix} + \frac{\partial A}{\partial \xi_n}e^{i\theta_n}
\begin{Bmatrix}
-2im_1\omega c_g \phi_u-K_2\phi_ve^{-ik}\\ -2im_2\omega c_g \phi_v+K_2\phi_ue^{ik} 
\end{Bmatrix}  \nonumber \\
	&+ K_3A^2e^{i2\theta_n}\begin{Bmatrix}
(1-e^{-i2k})\phi_v^2-2(1-e^{-ik})\phi_v\phi_u \\
(e^{i2k}-1)\phi_u^2-2(e^{ik}-1)\phi_v\phi_u
\end{Bmatrix} + c.c,
\label{eqf1}
\end{align}
where $\boldsymbol{\phi} = \begin{Bmatrix} \phi_u, \; \phi_v \end{Bmatrix}^T$, and $c.c$ refers to the complex conjugate of all the terms listed.
Due to the linearity of the equations in the multiple scales expansion, the forced (particular) solution at $\mathcal{O}(\varepsilon)$ can be represented as a superposition of the response due to the three independent terms in eqn.~(\ref{eqf1}). In this regard, if any of the terms in the forcing function resonate with the homogeneous solution (feature the same functional dependence on $k, \omega$ or $e^{i\theta_n}$), then the forced solution corresponding to these terms become unbounded (grow with time $t$, rather than being oscillatory), which violates the assumption of the perturbation expansion. Such terms in the forcing function are referred to as secular terms and need to be eliminated \cite{Huang1993, Huang1998}. The possible eigen solutions $(k, \omega, \boldsymbol{\phi})$ with which the terms in the forcing function $\bff^1$ can resonate are
\begin{align}
[\omega, k] &=  [0,0], \,\left[\sqrt{\frac{K_2}{m_1m_2}(m_1+m_2) - \frac{K_2}{m_1m_2} \sqrt{m_1^2 + m_2^2 + 2m_1m_2\cos{k}}},\,k\right], \nonumber\\
\boldsymbol{\phi} &= \begin{Bmatrix} C \\ C \end{Bmatrix} , \begin{Bmatrix}
\frac{-(1+e^{-ik})m_2}{m1-m2-\sqrt{m_1^2 + m_2^2 + 2m_1m_2\cos{k}}}\\ 1 \end{Bmatrix}, \label{eqeigmod}
\end{align}
where $C$ is a constant. When the forcing function has the same functional form as the eigen solution, it is convenient to employ a modal decomposition to obtain the forced solution \cite{meirovitch2001}. Specifically, in order to eliminate secular solutions, we require the corresponding forcing function terms to vanish in the modal coordinates \cite{Manktelow2014}, i.e.,
\begin{equation}
\boldsymbol{\phi}^H\bff=0,
\end{equation}
where $\boldsymbol{\phi}^H$ refers to the complex conjugate (or Hermitian) transpose of the eigenmode $\boldsymbol{\phi}$. Considering the eigenmodes in eqn.~(\ref{eqeigmod}) and the corresponding forcing function terms, it can be shown that 
\begin{align}
\begin{Bmatrix} C \\ C \end{Bmatrix}^H & \begin{Bmatrix}
-K_2 \frac{\partial A_{0}}{\partial \xi_n} \\ K_2 \frac{\partial A_{0}}{\partial \xi_n} 
\end{Bmatrix} = 0, \label{eqp1sec1}\\
\begin{Bmatrix}
\frac{-(1+e^{-ik})m_2}{m1-m2-\sqrt{m_1^2 + m_2^2 + 2m_1m_2\cos{k}}}\\ 1 \end{Bmatrix}^H &\begin{Bmatrix}
-2im_1\omega c_g \phi_u-K_2\phi_ve^{-ik}\\ -2im_2\omega c_g \phi_v+K_2\phi_ue^{ik} 
\end{Bmatrix} = 0,   \label{eqp1sec2}
\end{align}
which implies that the orthogonality conditions are automatically satisfied, and there are no secular terms corresponding to the two eigen modes. While the orthogonality of the forcing function corresponding to the eigen-pair $(0,0)$ is trivially established, the orthogonality of the forcing function corresponding to $e^{i\theta_n}$ (eqn.~\ref{eqp1sec2}) is obtained through the particular choice of velocity associated with the slow spatiotemporal coordinate $\xi_n$. In general, it is customary to leave the velocity of the slow spatiotemporal scale $\xi_n$ undetermined, and choose an appropriate value such that the secular terms vanish \cite{Huang1993}. In this case, choosing a priori this quantity to be the group velocity of the linear wave $(c_g)$ automatically nullifies the secular terms corresponding to $e^{i\theta_n}$. Therefore, the forced solution to eqn.~(\ref{eq4}) at $\mathcal{O}(\varepsilon)$ is only due to the term proportional to $e^{i2\theta_n}$ in eqn.~(\ref{eqf1}). As a result, the forced solution will also be of the form $e^{i2\theta_n}$, which implies that a second harmonic at $2\omega$ is generated due to the presence of nonlinearity in the system. Now, under the assumption that the forcing function corresponding to $e^{i2\theta_n}$ does not resonate with the homogeneous solution (i.e., $\omega(2k) \neq 2\omega(k)$), the particular (forced) solution at $\mathcal{O}(\varepsilon)$ can be written as\footnote{The negation of this condition is referred to as phase matching or resonance of the generated second harmonic. Under such conditions, additional constraints need to be incorporated to determine the forced solution (similar to the group velocity constraint imposed for the fundamental harmonic). Typically, a three-wave-interaction model is assumed to study resonant second harmonic generation \cite{Konotop1996,Goncalves2000}, which will not be considered in this work.}
\begin{equation}
\bfu_n^{1p} = B(\xi_n,\tau)\,\boldsymbol{\Phi}\,e^{i2\theta_n}+c.c, 
\label{eq4.2om}
\end{equation}
where  $B=K_3A^2$, and 
\begin{equation}
\boldsymbol{\Phi} = \begin{bmatrix}
-4\omega^2m_1+2K_2 & -K_2(1+e^{-i2k}) \\ 
-K_2(1+e^{i2k}) & -4\omega^2m_2+2K_2
\end{bmatrix}^{-1}\begin{Bmatrix}
(1-e^{-i2k})\phi_v^2-2(1-e^{-ik})\phi_v\phi_u \\
(e^{i2k}-1)\phi_u^2-2(e^{ik}-1)\phi_v\phi_u 
\end{Bmatrix}. \label{eq4.2omp}
\end{equation}
\indent The homogeneous component of the solution at $\mathcal{O}(\varepsilon)$ is chosen such that the initial conditions specified for the diatomic chain are satisfied. For a semi-infinite chain excited at one end with a harmonic excitation, the initial condition at $\mathcal{O}(\varepsilon)$ corresponds to imposing the absence of second harmonic features at the first mass \cite{Cabaret2012,Morcillo2013}. Therefore, the homogeneous solution at $\mathcal{O}(\varepsilon^1)$ can be written as
\begin{equation}
\bfu_n^{1h} = D(\xi_n,\tau)\,\boldsymbol{\chi}\,e^{i\psi_n}+ c.c,
\end{equation}
where $\psi_n = k(2\omega)n-2\omega t$ is the fast spatiotemporal variable that satisfies the linear eigenvalue problem. $\boldsymbol{\chi}$ is the eigenvector corresponding to the eigenfrequency pair $[2\omega,k(2\omega)]$, and $D$ is determined from the condition $u_0^1=0$. In conclusion, the complete solution at $\mathcal{O}(\varepsilon^1)$ can be written as 
\begin{equation}
\bfu_n^{1} = D(\xi_n,\tau)\,\boldsymbol{\chi}\,e^{i\psi_n} + B(\xi_n,\tau)\,\boldsymbol{\Phi}\,e^{i2\theta_n} + c.c,	
	\label{eq4.2p1}
\end{equation}
\indent Now, assuming that the equations up to $\mathcal{O}(\varepsilon)$ effectively capture the response of the nonlinear system modeled by eqn.~(\ref{eq4.1}), the general solution can be obtained by combining eqns.~(\ref{eqrefp0}) and (\ref{eq4.2p1}) as
\begin{equation}
\bfu_n =  A_0(\xi_n,\tau) + A(\xi_n,\tau)\,\boldsymbol{\phi}\,e^{i\theta_n}+ \varepsilon D(\xi_n,\tau)\,\boldsymbol{\chi}\,e^{i\psi_n} + \varepsilon B(\xi_n,\tau)\,\boldsymbol{\Phi}\,e^{i2\theta_n} + c.c.
\label{eq4.13}
\end{equation}
\indent The exact solutions for $A_0$ and $A$ (and subsequently $B$ and $D$) have not been determined; this task requires extending the analysis to equations at $\mathcal{O}(\varepsilon^2)$ and higher. Following the same procedure employed at $\mathcal{O}(\varepsilon)$, the suppression of secular terms yields the Nonlinear Schr\"odinger equation, which governs the dependence of $A_0$ and $A$ as a function of $\xi$ and $\tau$. Since these expressions are already available in published literature \cite{Pnevmatikos1986,Huang1998}, we will not report them here for the sake of brevity. Instead, we will place our focus on the functional structure of the solution in eqn.~(\ref{eq4.13}), and use it as a conceptual platform to delineate and interpret the peculiar characteristics of the nonlinearly-generated harmonics, which we will exploit later in physically realizable lattice structures.\newline
\indent Let us now systematically interpret the terms of eqn.~(\ref{eq4.13}), which represent different coexisting contributions of a general wave response. The solution at the leading order ($\mathcal{O}(1)$) consists of a fundamental harmonic term modulated by a slow scale envelope $A(\xi_n,\tau)$ and an additional constant slow scale envelope term $A_0(\xi_n,\tau)$. In single degree-of-freedom, dispersive, linear systems, it can be shown that the constant term $A_0(\xi_n,\tau)$ vanishes, and the expression for $A(\xi_n,\tau)$ is obtained by solving the linear Schr\"odinger equation \cite{Remoissenet1999}. In the case of weakly nonlinear systems, the constant envelope term depends on the magnitude of $A(\xi_n,\tau)$ and the  magnitude of quadratic nonlinearity $(K_3)$ in the system \cite{Huang1993}. In essence, the effect of nonlinearity on the leading order solution manifests as a long-wavelength modulation of the response of the corresponding linear chain \cite{GG13}. We have analyzed the structure of the long-wavelength modulation in nonlinear diatomic chains in \cite{GG14}, and shown that it can be parameterized as a function of the nonlinearity and the excitation parameters, and can be used to setup inverse problems to determine nonlinear coefficients from numerical or experimental data. In summary, the envelope modulation is an integral feature associated with the dispersion which is also affected by the presence of nonlinearity in the system.\newline
\indent In contrast, the higher-order terms ($\mathcal{O}(\varepsilon)$) in the solution  are completely attributed to the nonlinear terms and reflect the well-known generation of harmonics. While the nonlinearly-generated harmonics also feature an envelope modulation (given by the terms $B(\xi_n,\tau), D(\xi_n,\tau)$ here), the most interesting aspect capturing the interplay of dispersion and nonlinearity is the presence of modal complexity (denoted by the modal vectors $\boldsymbol{\Phi}$ and $\boldsymbol{\chi}$, which effectively describe the relative motion of the two masses in the unit cell). Specifically, the second harmonic has multiple, fast spatiotemporal characteristics (associated with different modal contribution), one corresponding to the forcing term $(2\theta_n=2k(\omega)n-2\omega t)$, and the other corresponding to the homogeneous solution $(\psi_n=k(2\omega)n-2\omega t)$. In monoatomic periodic structures, these two components can be viewed as a single second harmonic feature modulated by a spatially beating amplitude $(B\propto e^{i(2k-k(2\omega))n})$ \cite{Bradley1995,Morcillo2013}. On the contrary, in periodic structures characterized by multiple modes of wave propagation, the two second harmonic components have different modal distributions ($\boldsymbol{\Phi}$ v.s. $\boldsymbol{\chi}$), thereby not necessarily giving rise to spatial beating. More importantly, the modal contribution $\boldsymbol{\chi}$ of the homogeneous part of the nonlinearly-generated second harmonic conforms to the dispersion relation of the corresponding linear structure, which implies that the characteristics of this component can be determined solely from a simple eigen analysis of the corresponding linear structure. On the other hand, the modal contribution $\boldsymbol{\Phi}$ of the forced second harmonic is determined from eqn.~(\ref{eq4.2omp}) and depends on the characteristics of the fundamental mode. Therefore, the fact that $\boldsymbol{\chi}$ can be predicted by simply considering a linear system allows us to potentially design the modal characteristics of this component through a simple unit cell analysis. In order to further elaborate on this statement and to visually explain the intricate effects of modal complexity (and the resulting opportunities for tunability and wave control), we will resort to a numerical simulation of the diatomic spring-mass system.
\subsection{Modal mixing in nonlinear diatomic chains}
\begin{figure}[!htb]
\centering
\subfloat[Spatiotemporal response for $\alpha=0.3$]{\label{f4.2a}\includegraphics[scale=0.45]{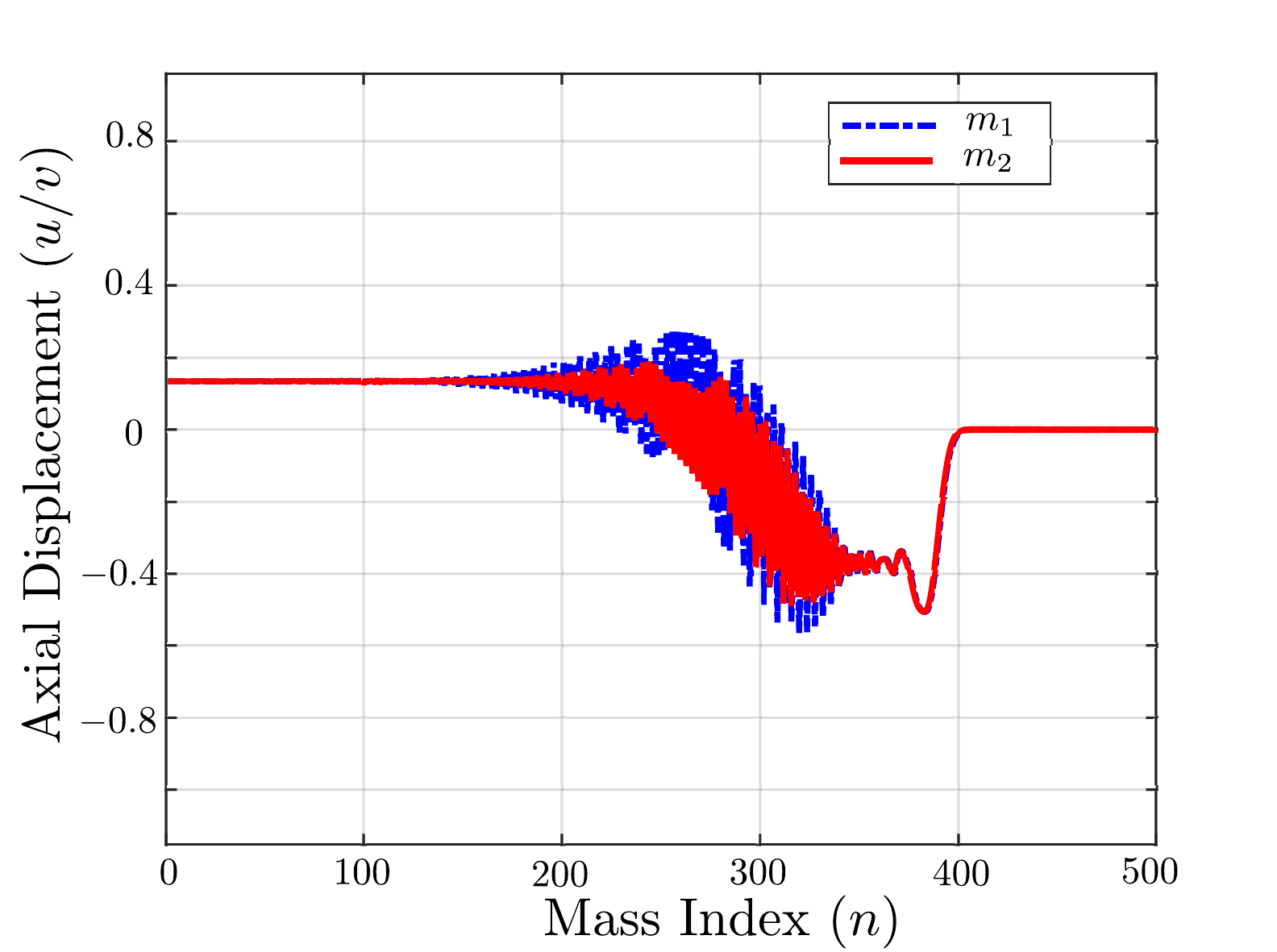}} \qquad
\subfloat[Spatiotemporal response for $\alpha=0.5$]{\label{f4.2b}\includegraphics[scale=0.45]{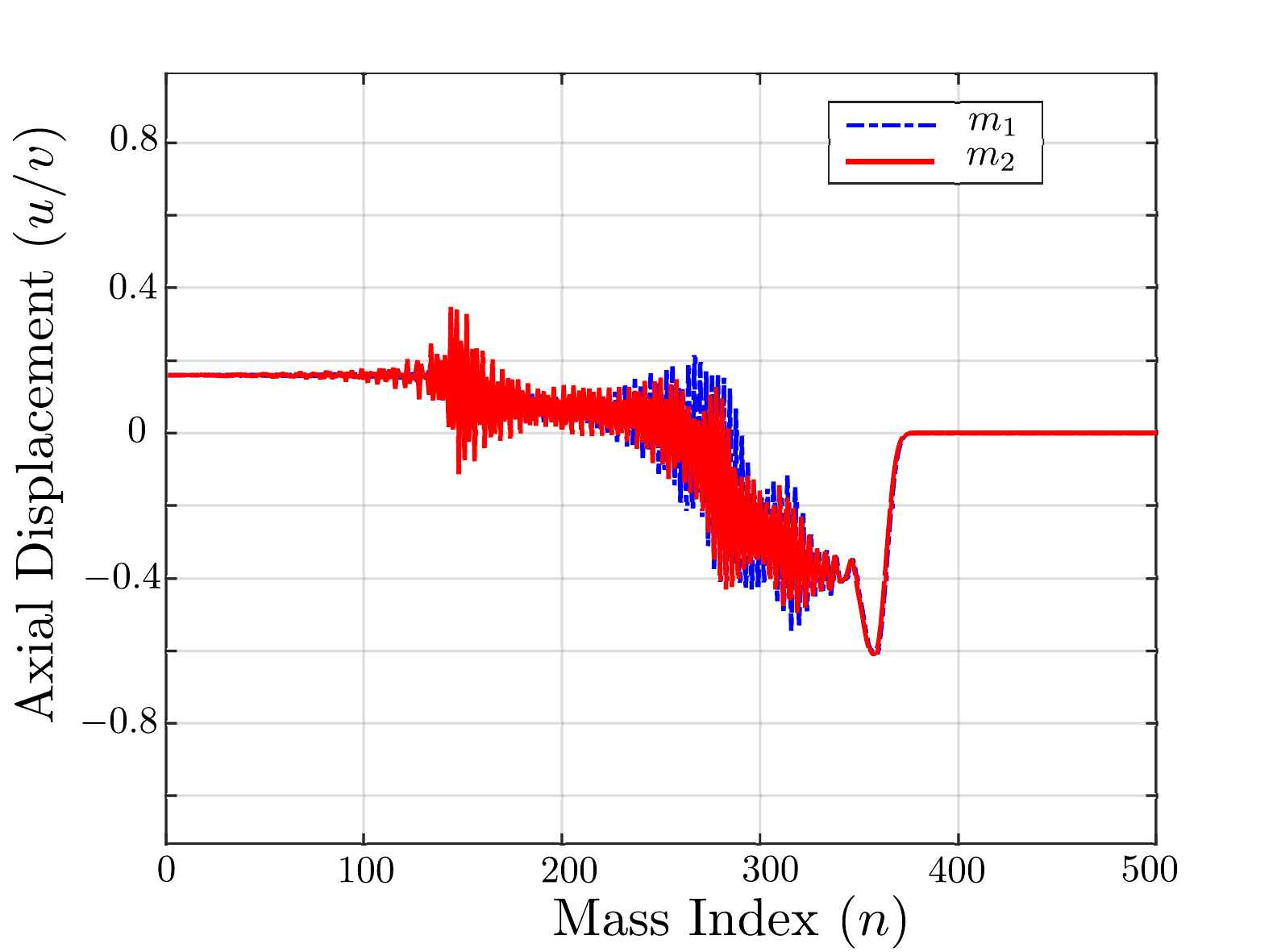}} \\
\subfloat[Spectral response for $\alpha=0.3$]{\label{f4.2c}\includegraphics[scale=0.45]{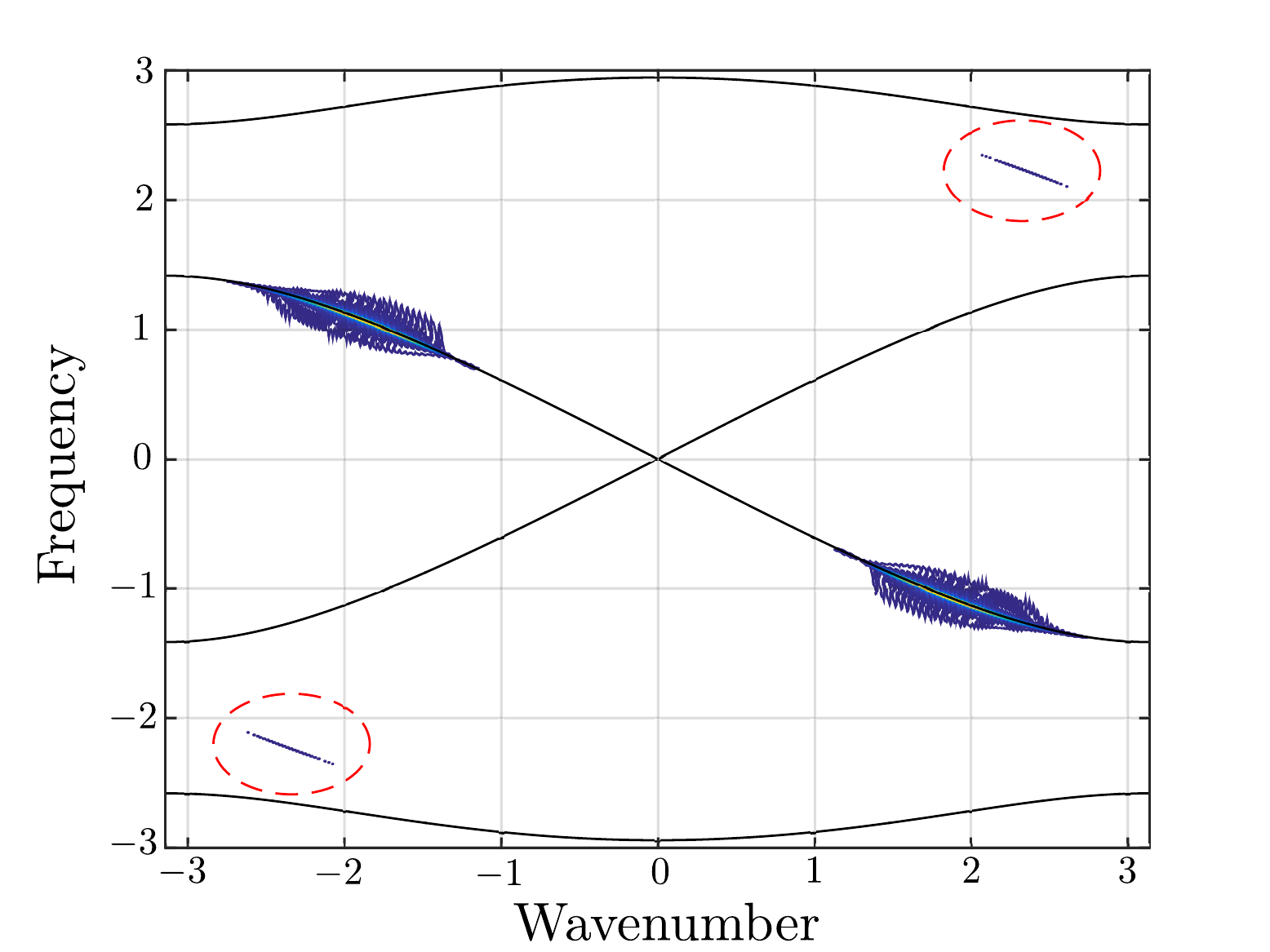}} \qquad
\subfloat[Spectral response for $\alpha=0.5$]{\label{f4.2d}\includegraphics[scale=0.45]{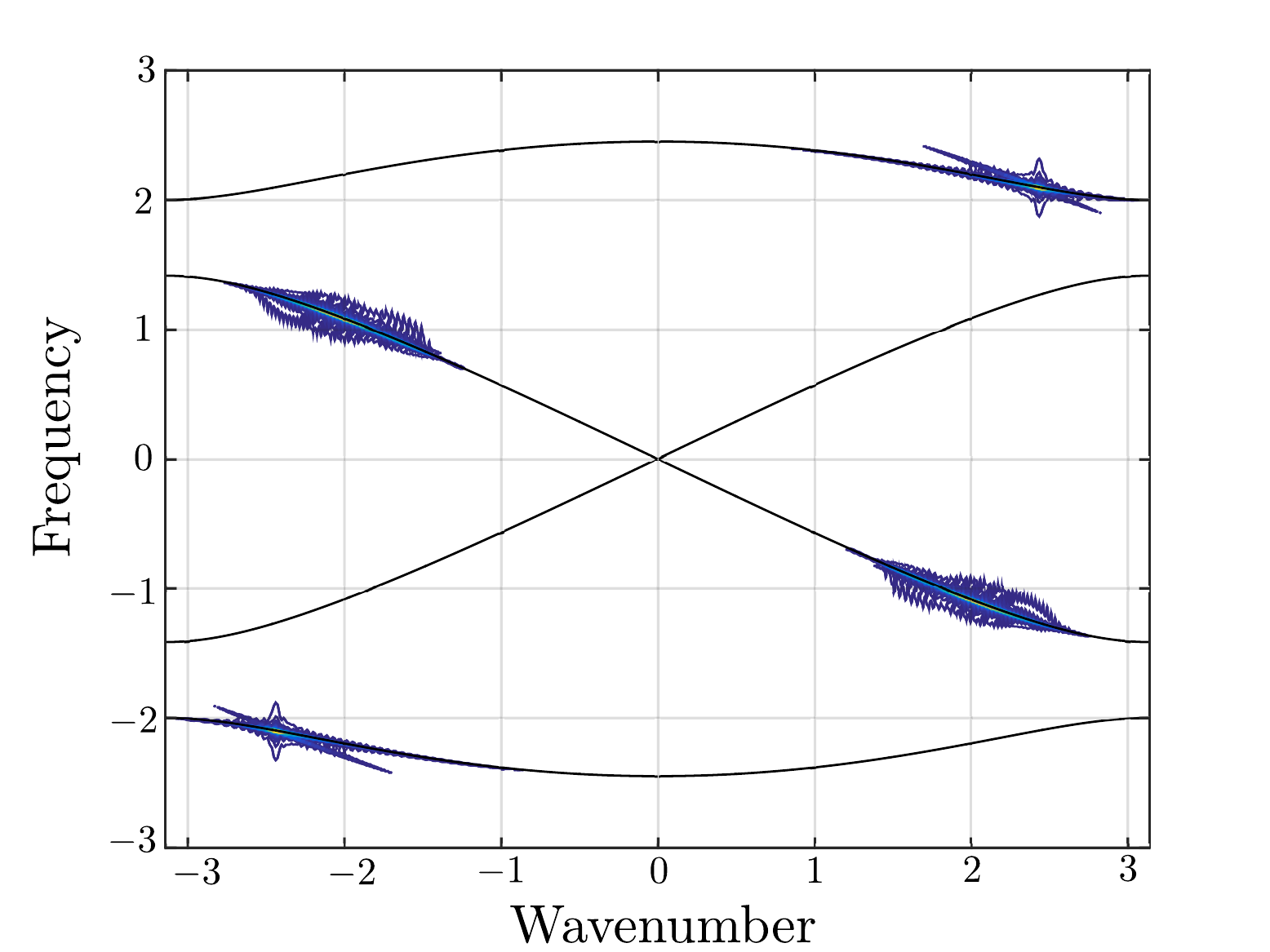}} 
\caption{Spatial and spectral wave profile in two diatomic chains characterized by different mass ratios for the same input excitation (a-b) Spatial wave profile, (c-d) $2$D DFT of the response superimposed to the band diagram evaluated from unit cell analysis.}
\label{fig4.2}
\end{figure}

We consider the numerical analysis of a semi-infinite diatomic spring-mass chain (with $m_1=1,m_2=\alpha$; $\alpha<1$ for convenience) with unit linear and nonlinear spring coefficient $(K_2=K_3=1,\,K_4=0)$. A burst excitation with the carrier frequency chosen to belong to the acoustic mode of wave propagation is applied to the first mass and the spatiotemporal displacements (of all the masses) are computed by numerically integrating the equations of motion. The spatial wave profile (at a particular time instant) is plotted for two different choices of mass ratio $(\alpha = 0.3, 0.5)$ in figs.~\ref{fig4.2}(a-b), where we observe the presence of two distinct features - a non-harmonic and a harmonic component. The non-harmonic component is attributed to the slow scale envelope term $A_0(\xi_n,\tau)$ in the multiple scales expansion \cite{Huang1993,Pnevmatikos1986,GG14}.\newline 
\indent Focusing on the harmonic components of the spatial wave profile in figs.~\ref{fig4.2}(a-b), it can be noticed that, for the same excitation parameters, the spatial profile is significantly sensitive to changes in mass ratio. In order to understand the differences between these two spatial profiles, we compute the spectral contours of the response using a Discrete Fourier Transform (DFT), and the contours corresponding to the lighter mass $(\mathcal{F}(v_n))$ are shown for both cases in figs.~\ref{fig4.2}(c-d). The long-wavelength envelope, which dominates the spectral response has been filtered out to highlight the harmonic components of the wave packets, and the contours are superimposed to the dispersion branches computed from unit cell analysis\footnote{The cutoff frequency of the acoustic mode is independent of the mass contrast $(\alpha)$, while the position of the optical mode and, consequently, the width of the bandgap is significantly affected by the mass contrast.}. Here, the spectral contours are plotted in the entire spectral domain ($\pm k, \pm \omega$) as the acoustic and optical mode for the diatomic chain have complementary wave characteristics (i.e., the group velocity of the acoustic and the optical wave packets have opposite signs), which implies that they occupy different quadrants in the spectral domain.\newline
\indent For $\alpha=0.3$, the nonlinearly-generated second harmonic $2\omega$ lies in the bandgap, i.e., there exists no real value of $k$ that satisfies the dispersion relation at $2\omega$. Therefore, the homogeneous component has attenuating characteristics and only the second harmonic component corresponding to the forced solution ($2kn - 2\omega t$) propagates through the structure. The spectral component corresponding to $2kn-2\omega t$ appears in the opposite quadrant of wave propagation (as highlighted in fig.~\ref{fig4.2}c) due to the fact that the resolution of the wavenumber in the band diagram is only limited to the fundamental period ($k \in [-\pi,\, \pi]$) \cite{Konotop1996}. In contrast, for $\alpha=0.5$, the second harmonic lies on the optical mode, which gives rise to a propagating characteristic for the homogeneous component of the nonlinearly-generated second harmonic. As a result, the amplitude of the second harmonic is much larger than that of the smaller mass ratio case. From these observations, we conclude that, when the choice of the fundamental frequency is such that the generated second harmonic lies in the bandgap of the periodic structure, the structure effectively acts as an adaptive filter, converting a part of the energy from the initial wave packet into a non-propagating mode. Conversely, a suitable choice of frequency such that the second harmonic lies in the domain of the optical mode leads to the phenomenon of \textit{mode hopping} \cite{GG15}, whereby a part of the energy injected into the system hops from the acoustic to the optical mode. Therefore, this component of the energy propagates through the system with fundamentally different modal characteristics than those of the applied excitation.\newline
\indent These considerations lead us to a few conceptual conclusions of paramount importance for metamaterials design. As observed from eqns.~(\ref{eq4.2omp}) and (\ref{eq4.2p1}), the nonlinearly generated harmonics encompass a homogeneous solution and a forced solution. Since the characteristics of the homogeneous component only depend on the properties of the linearized unit cell, they can be engineered to feature desired (e.g. complementary) modal characteristics by simply altering the parameters of the linearized structure, thereby providing an opportunity for adaptive spatial wave manipulation. In contrast, the forced component cannot be independently manipulated as the modal characteristics are dependent on the applied excitation. While the same concepts can be extended to the higher harmonics (third, fourth, and so on) by considering the equations at $\mathcal{O}(\varepsilon^2)$ and higher, we assume that these effects can be neglected due to the weak nature of nonlinearity in the system.\newline
\indent We have previously demonstrated the concept of mode hopping in granular phononic crystals, where the nonlinearity stems from the contact interaction between beads \cite{GG15}. However, the physical realization of granular structures poses many challenges, as effects such as disorder, plasticity, dissipation, and additional rotational degrees of freedom can affect the ideality of the system. Another route to realize nonlinear phononic crystals consists of employing lattice structures made of materials which can undergo large deformation. While the literature on lattice materials in the field of phononic crystals has been, for the most part, limited to the linear regime, recent interest in nonlinear periodic structures has grown in the soft materials community, particularly with regards to the objective of developing tunable lattice structures. In this regard, soft materials display nonlinear stress-strain relationships, a feature often referred to as material nonlinearity, which is exploited to obtain finite-deformation effects. On the other hand, nonlinear strain-displacement relationships, available even in materials exhibiting a linear stress-strain dependence, can also be exploited to design lattice structures capable of supporting finite-deformation effects. For example, a curved beam exhibits an intrinsically nonlinear strain-displacement relationship even for small values of stress and strain, a fact that can be utilized as the building block for lattice structures exhibiting finite-deformation effects. In the following section, we will illustrate the materials design implications of this discussion by considering two lattice materials featuring the ability to support finite deformations through a combination of material and geometric nonlinearity, and we will elucidate how mode hopping can be exploited for the design of adaptive phononic switches.
\section{Nonlinear Lattices}
The lattices under consideration are assumed to exhibit a nonlinear strain-displacement relationship, which can be expressed using the standard Lagrangian framework of continuum mechanics as 
\begin{equation}
\bfE =  \half\left[(\nabla \bfu) + (\nabla \bfu)^T + (\nabla \bfu)^T (\nabla \bfu)\right],
\end{equation}
where $\bfu$ are the displacements and $\bfE$ is the Green strain tensor. In addition, we assume that the lattice is made of a compressible isotropic hyperelastic material, whose strain energy function ($\mathcal{W}$) can be expressed in terms of the invariants $(\mathcal{I}_1, \mathcal{I}_2, \mathcal{I}_3)$ of the Cauchy-Green strain tensor $(\bfC = \bfI + 2\bfE)$ as   
\begin{equation}
\mathcal{W}\,=\mathcal{W}(\mathcal{I}_1,\mathcal{I}_2,\mathcal{I}_3).
\end{equation}
The second-Piola Kirchoff stress tensor $\bfS$ is then obtained from $\mathcal{W}$ as 
\begin{equation}
\bfS =  \gamma_0(\mathcal{I}_1,\mathcal{I}_2,\mathcal{I}_3) \bfI + \gamma_1(\mathcal{I}_1,\mathcal{I}_2,\mathcal{I}_3) \bfC + \gamma_2(\mathcal{I}_1,\mathcal{I}_2,\mathcal{I}_3) \bfC^2, 
\label{eqrefSHyp}
\end{equation}
where $\gamma_0,\,\gamma_1$, and $\gamma_2$ are determined from the derivatives of the energy function $\mathcal{W}$ with respect to the invariants $\mathcal{I}_1, \mathcal{I}_2, \mathcal{I}_3$. While eqn.~(\ref{eqrefSHyp}) represents a generic nonlinear stress-strain relation, it can be linearized for small displacements and expressed as a function of two constants (equivalent St. Venant-Kirchoff model) \cite{Jog_Continuum}. Therefore, assuming weak, but finite-displacements, and a linearized stress-strain relationship, the second Piola Kirchoff stress tensor can be simplified as 
\begin{equation}
\bfS(\bfE) = \overline{\lambda}\,tr(\bfE)\,\bfI + 2\,\overline{\mu}\,\bfE = \overline{\lambda}\,tr(\boldsymbol{\epsilon})\,\bfI + 2\,\overline{\mu}\,\boldsymbol{\epsilon} + \mathcal{O}(\varepsilon),
\end{equation}
where $\overline{\lambda} = f(\gamma_0,\gamma_1,\gamma_2)$, $\overline{\mu} = g(\gamma_1,\gamma_2)$, and $\boldsymbol{\epsilon} (= 1/2[(\nabla \bfu) + (\nabla \bfu)^T])$ is the small-strain tensor. $\mathcal{O}(\varepsilon)$ encompasses all the terms that have a quadratic dependence on the displacements. Using this expansion, the governing equations of motion can be expressed in terms of the displacement $\bfu$ as \cite{DeLima2003} 
\begin{equation}
(\overline{\lambda}+2\overline{\mu})\,\nabla(\nabla\cdot\bf u) - \overline{\mu}\,\nabla\times(\nabla\times\bf u) + \varepsilon \bfF_{nl} = \rho_0 \ddot{\bf u},
\end{equation}
where $\rho_0$ is the density of the material in the reference configuration, and $\bfF_{nl}$ contains the nonlinear terms that have a quadratic dependence on the displacements due to the strain-displacement relationship. In the case of cellular solids, the lack of analytical solutions forces one to employ numerical techniques such as the finite element method. Therefore, the governing differential equation can be recast in discretized form as 
\begin{equation}
\textbf{M} \ddot{\textbf{u}} + \textbf{Ku} + \varepsilon \textbf{f}_{NL} (\textbf{u}) = 0,
\label{eqref}
\end{equation}
where $\bfM$ and $\bfK$ are the Mass and linearized Stiffness matrices, and $\bfu$ is the vector of nodal displacements. $\textbf{f}_{NL}(\textbf{u})$ is the vector that represents the nonlinear terms. While we have only incorporated the effect of geometric nonlinearity in deriving eqn.~(\ref{eqref}), the presence of material nonlinearity (due to the nonlinear stress-strain relationship expressed in eqn.~\ref{eqrefSHyp}) would only augment the nonlinear force term $\textbf{f}_{NL}$ as the leading order dependence of the stress-strain relationship is captured by the linear model. Therefore, eqn.~(\ref{eqref}) also represents the general discretized model valid for a structure that exhibits material nonlinearity. By reducing eqn.~(\ref{eqref}) to that of the unit cell, it can be easily shown that the equations are similar to eqn.~(\ref{eq4.1}). Therefore, the observations made in section \ref{sec2} regarding the morphological characteristics of nonlinear traveling wave packets can be extended to generic structures governed by eqn.~(\ref{eqref}), without loss of generality. \newline
\indent In order to illustrate the multi-modal characteristics of finite-amplitude waves in lattice structures that can be represented by eqn.~(\ref{eqref}), we will resort to the full-scale numerical analysis of lattices with different geometric and material properties, and we will demonstrate that it is sufficient to know the characteristics of the linearized unit cell in order to design the activation of adaptive and complementary modal functionalities. First, we will consider an example where the nonlinearity is established solely due to geometric mechanisms, and then extend the analysis to structures featuring material nonlinearity.
\subsection{Curved Lattice}
\begin{figure}[!htb]
\centering
\includegraphics[scale=0.2]{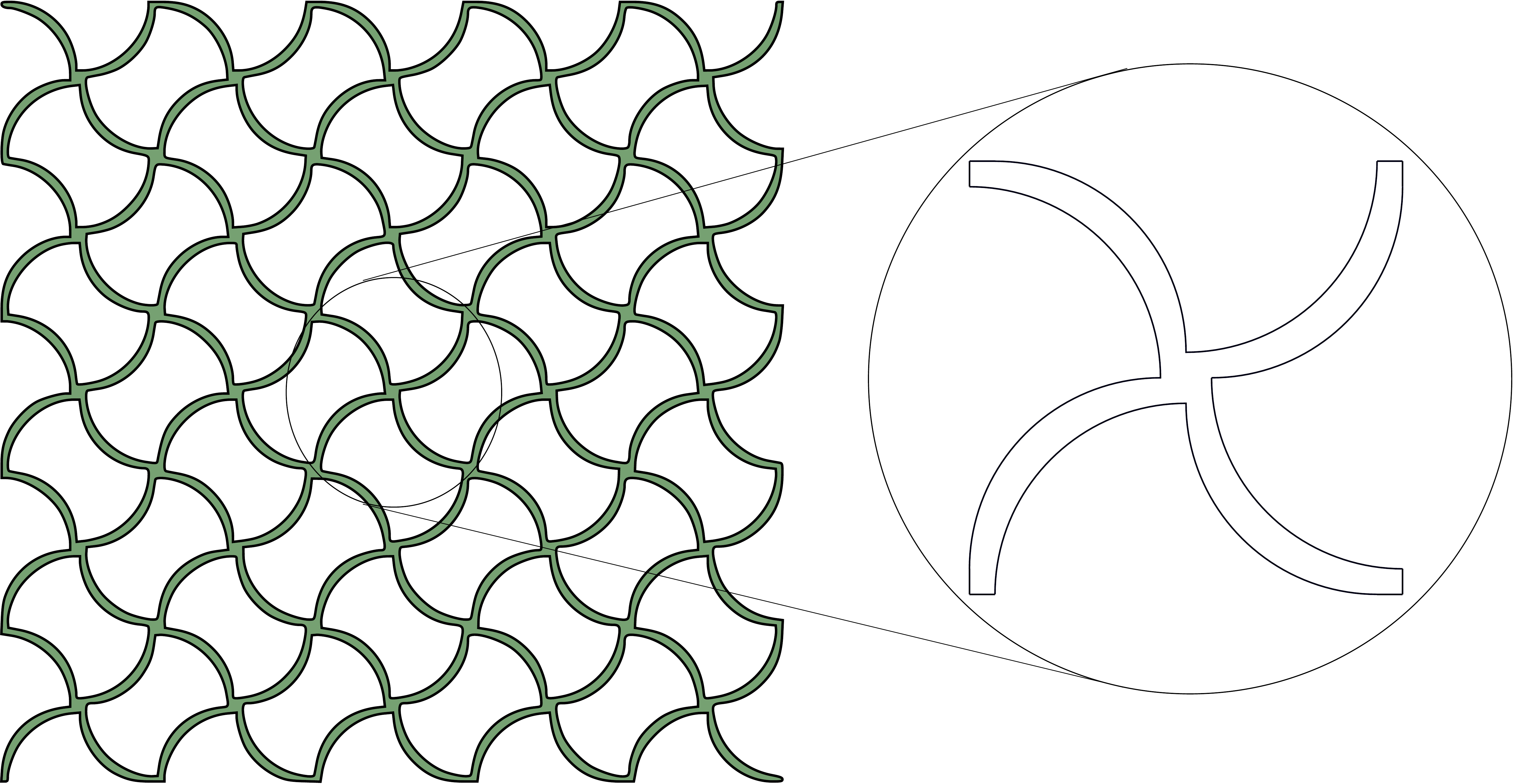}
\caption{Schematic of the four-blade propeller lattice with the unit cell highlighted in the inset.}
\label{fig52.1}
\end{figure}

The lattice under consideration consists of a four-blade propeller shaped unit cell tessellated along the horizontal and vertical directions, as shown in fig.~\ref{fig52.1}. The curvature of the links imparts them the ability to accommodate large rotations, which manifests as a nonlinear strain component, and allows us to realize the nonlinear features elaborated in section \ref{sec2}. 
For this lattice, we assume that the stress-strain relationship is linear and can be expressed using a St. Venant-Kirchoff model as 
\begin{equation}
\label{eqCurvS}
\bfS = \lambda\,tr(\bfE)\,\bfI + 2\,\mu\,\bfE,
\end{equation}
where $\lambda$ and $\mu$ are the Lam\'e parameters. We choose a set of test parameters: $\lambda = 2000$, $\mu= 1000$, and $\rho_0=1$. In order to determine the spatial characteristics of wave propagation, we first analyze the corresponding linearized unit cell (obtained by replacing $\bfE$ in eqn.~(\ref{eqCurvS}) with $\boldsymbol{\epsilon}$), modeled using 2D elasticity and discretized using isoparametric 4-node plane stress elements. \newline 
\begin{figure}[!htb]
\vspace{0.05 in}
\centering
\includegraphics[scale=0.62]{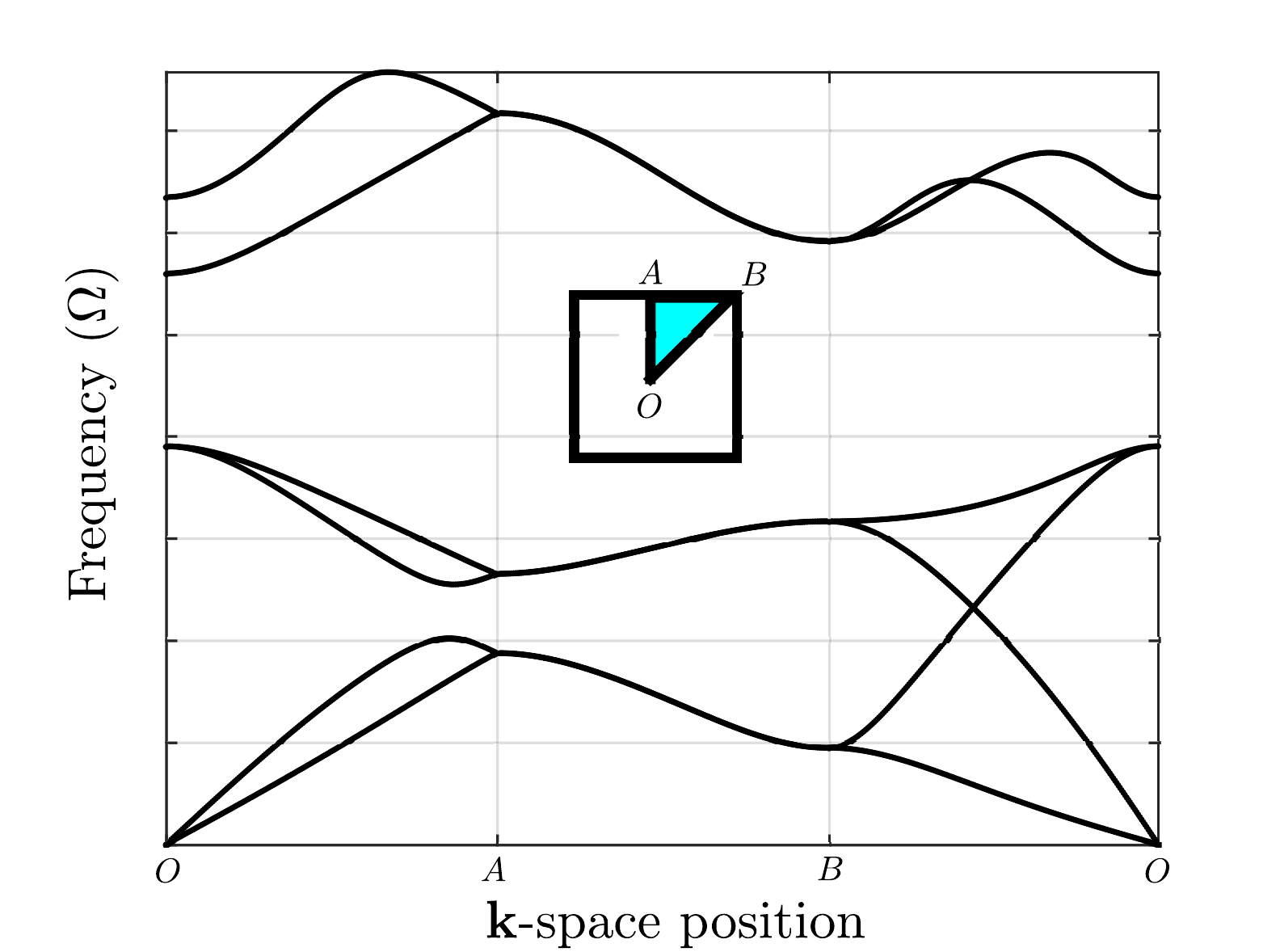} 
\caption{Linearized band diagram of the curved lattice modeled using 2D plane stress elasticity to capture in-plane wave propagation. The band structure is sampled along the contour of the irreducible Brillouin zone $OABO$ (shown in the inset).}
\label{fig52.3}
\end{figure}
\indent The dispersion relation $\omega(\bf k)$ for in-plane wave propagation ($\textbf{k} = k_x\hat{i} + k_y\hat{j}$) is estimated by following a standard procedure in phononics analysis \cite{Phani2006}, whereby Floquet-Bloch conditions are applied at the boundary of the unit cell yielding a reduced eigenvalue problem in terms of reduced $\textbf{k}$-dependent mass and stiffness matrices. The band structure, shown in fig.~\ref{fig52.3}, is then obtained by solving the reduced eigenvalue problem for wavevectors sampled along the contour of the irreducible Brillouin zone. This representation provides a map for the adaptive functional augmentation mechanisms that can be achieved by mode hopping.\newline
\indent First, we observe the existence of a complete bandgap between the third-fourth and fifth-sixth modes of wave propagation, whose location and width are controlled by the geometric and material properties of the unit cell. As already discussed for the diatomic chain in section \ref{sec2}, for choices of frequencies such that the nonlinearly-generated harmonic falls within this bandgap, a part of the energy associated with the fundamental excitation is converted into a non-propagating mode. As a result, the parameters of the unit cell can be designed to switch on and off the ability of the lattice to selectively propagate part of energy associated with the fundamental excitation.\newline 
\indent An even more interesting feature of the band structure is the presence of a smaller, partial bandgap between the first-second acoustic and the third-fourth optical modes along $OA$ (which identifies waves propagating along the horizontal or vertical directions in the lattice). By establishing branch hopping condition across this bandgap through nonlinear harmonic generation (by controlling the amplitude of excitation), we can deploy some of the energy from an acoustic mode to an optical mode and thus induce spatial characteristics associated with the higher frequency modes, while exciting the structure at lower frequencies of excitation (in the range of the first two modes). This capability is especially interesting when the directivity landscapes associated with the modes feature a high degree of complementarity.
\begin{figure}[!htb]
\centering
\includegraphics[scale=0.28]{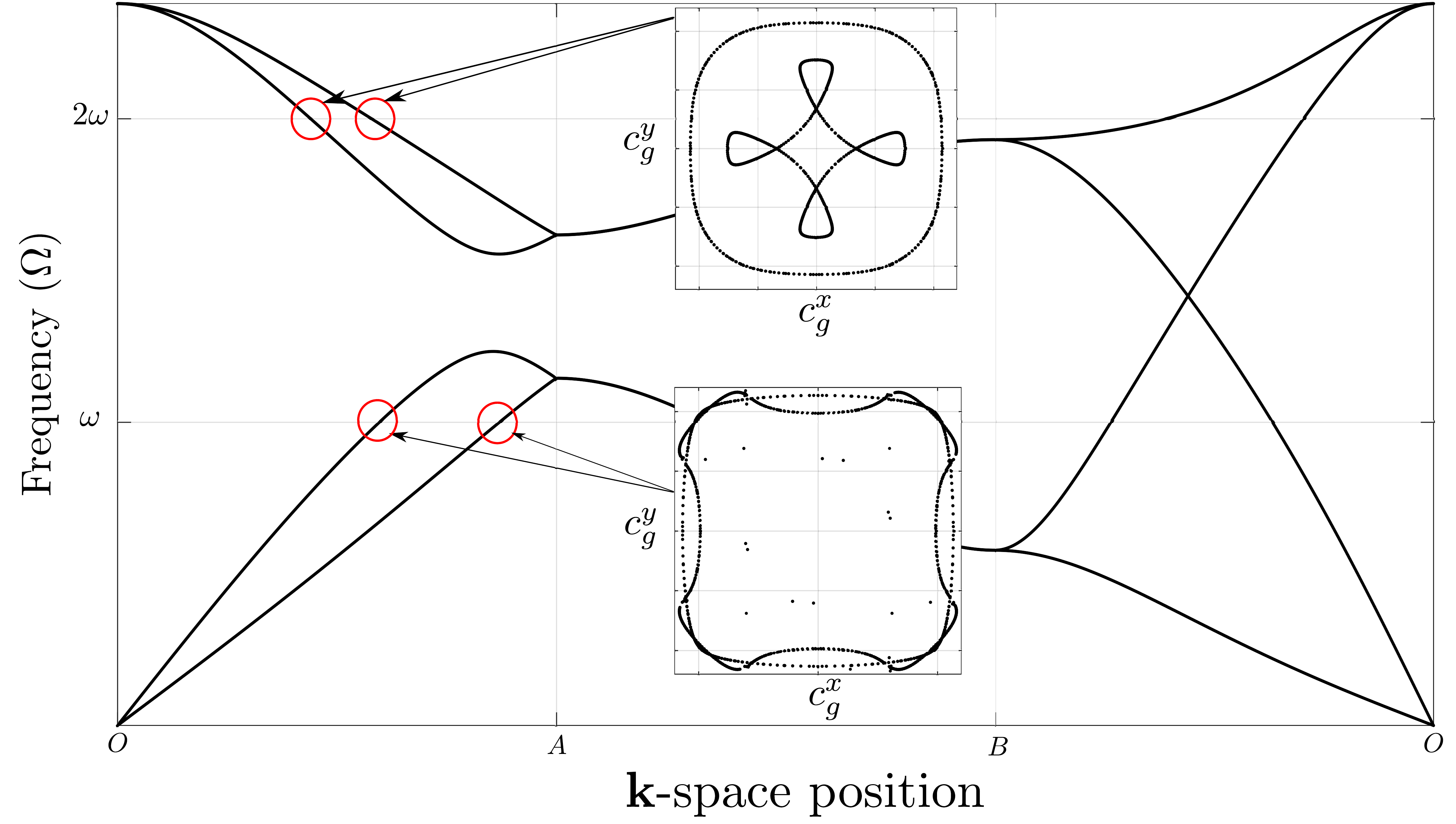}
\caption{Potential for directivity augmentation via mode hopping in a curved lattice. The group velocity contours at the chosen frequencies (shown in the inset) illustrate the complementary directional characteristics of acoustic and optical modes.}
\label{fig52.3b}
\end{figure}
We illustrate this capability by considering two frequencies ($\omega$ and $2\omega$) that intersect the branches of the dispersion curve as shown in fig.~\ref{fig52.3b}. The spatial characteristics of waves propagating at these frequencies are described by the group velocity contours (shown in the two insets of fig.~\ref{fig52.3b}), which are computed by evaluating the gradient of the dispersion surfaces, sampled at the values of wavevectors corresponding to the isofrequency contours at $\omega$ and $2\omega$, respectively.\newline
\begin{figure}[!htb]
\centering
\subfloat[Divergence of the wavefield excited at $\omega$]{\includegraphics[scale=0.25]{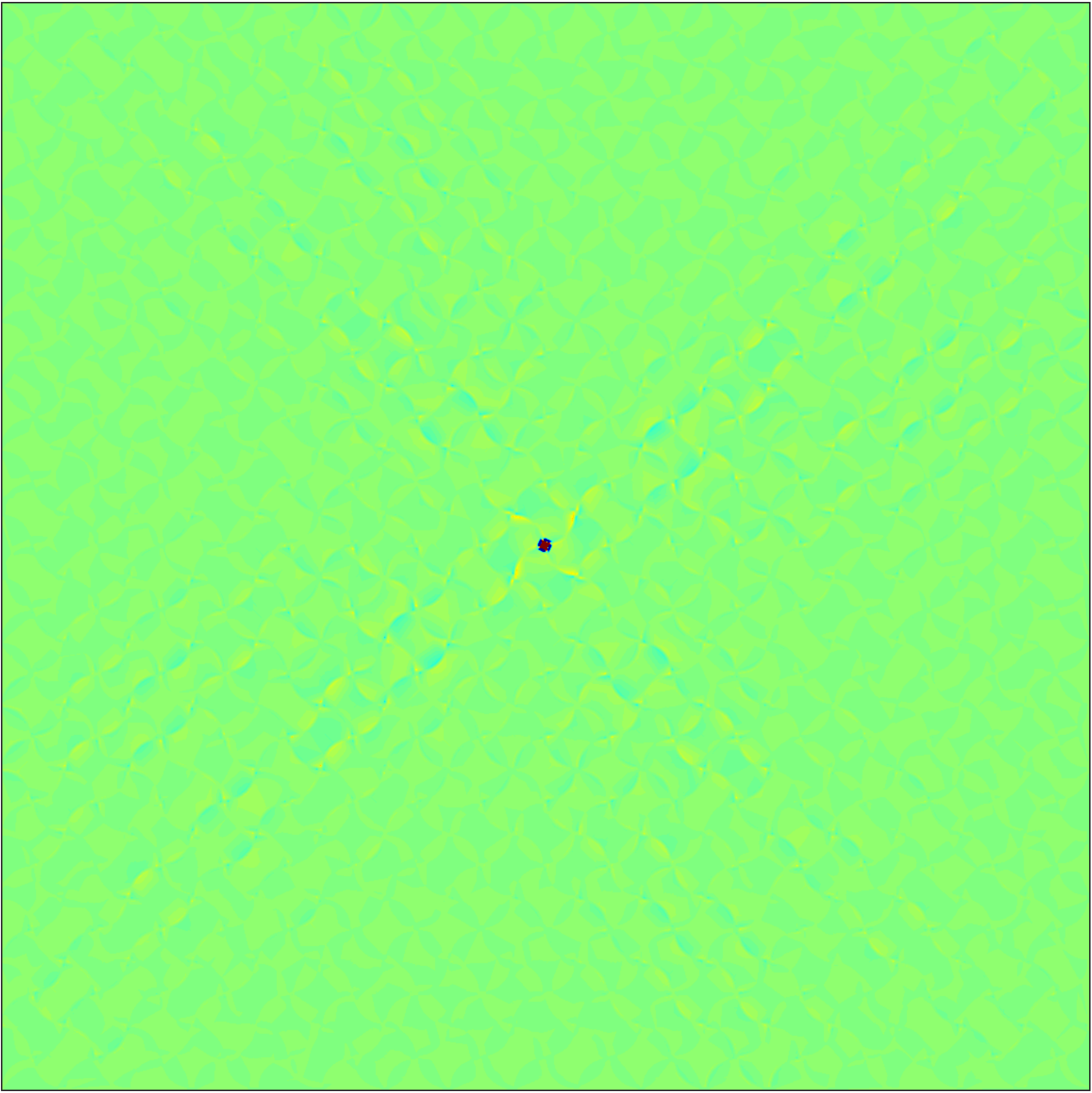}} \qquad\qquad
\subfloat[Curl of the wavefield excited at $\omega$]{\includegraphics[scale=0.25]{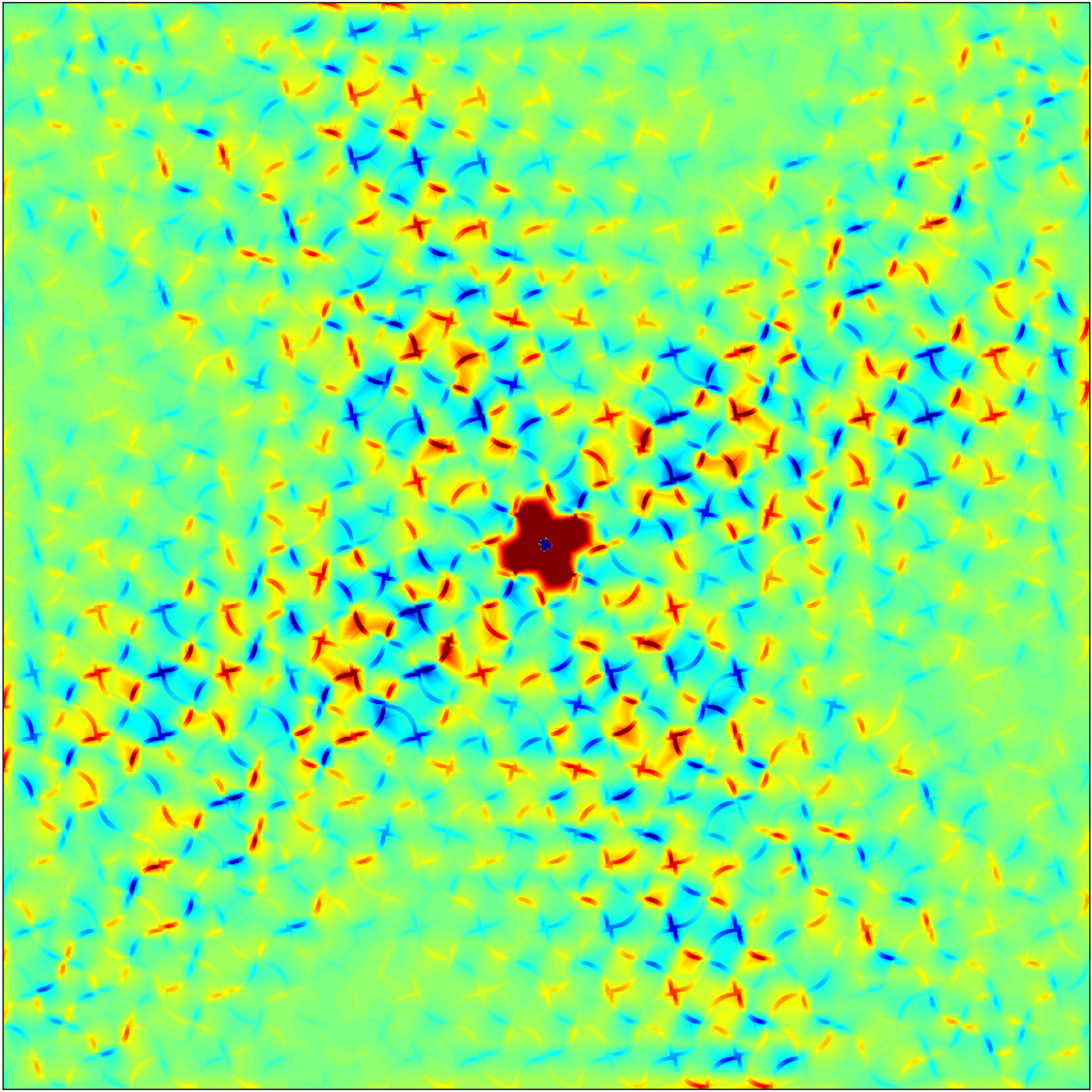}} \\
\subfloat[Divergence of the wavefield excited at $2\omega$]{\includegraphics[scale=0.25]{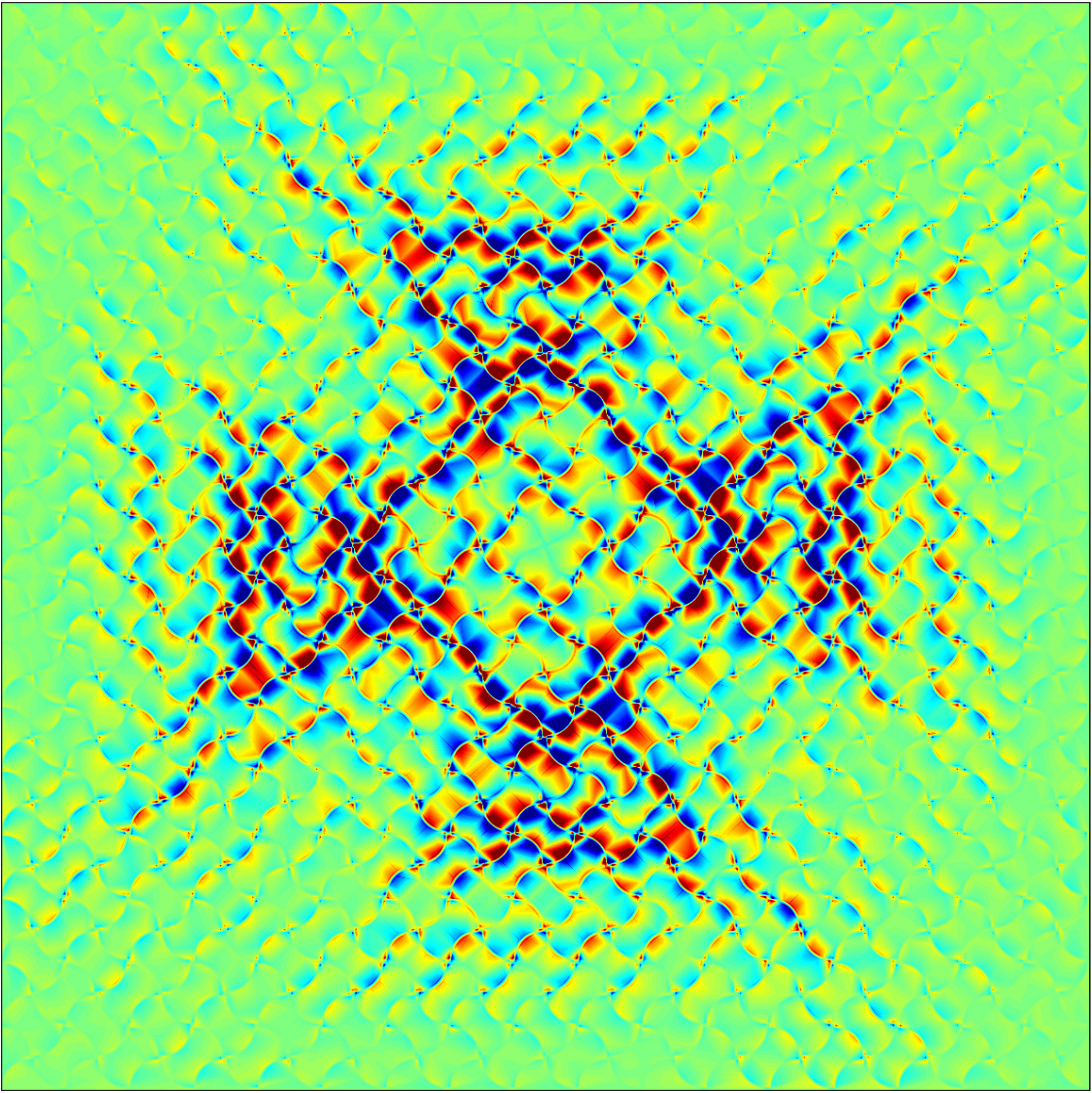}} \qquad\qquad
\subfloat[Curl of the wavefield excited at $2\omega$]{\includegraphics[scale=0.25]{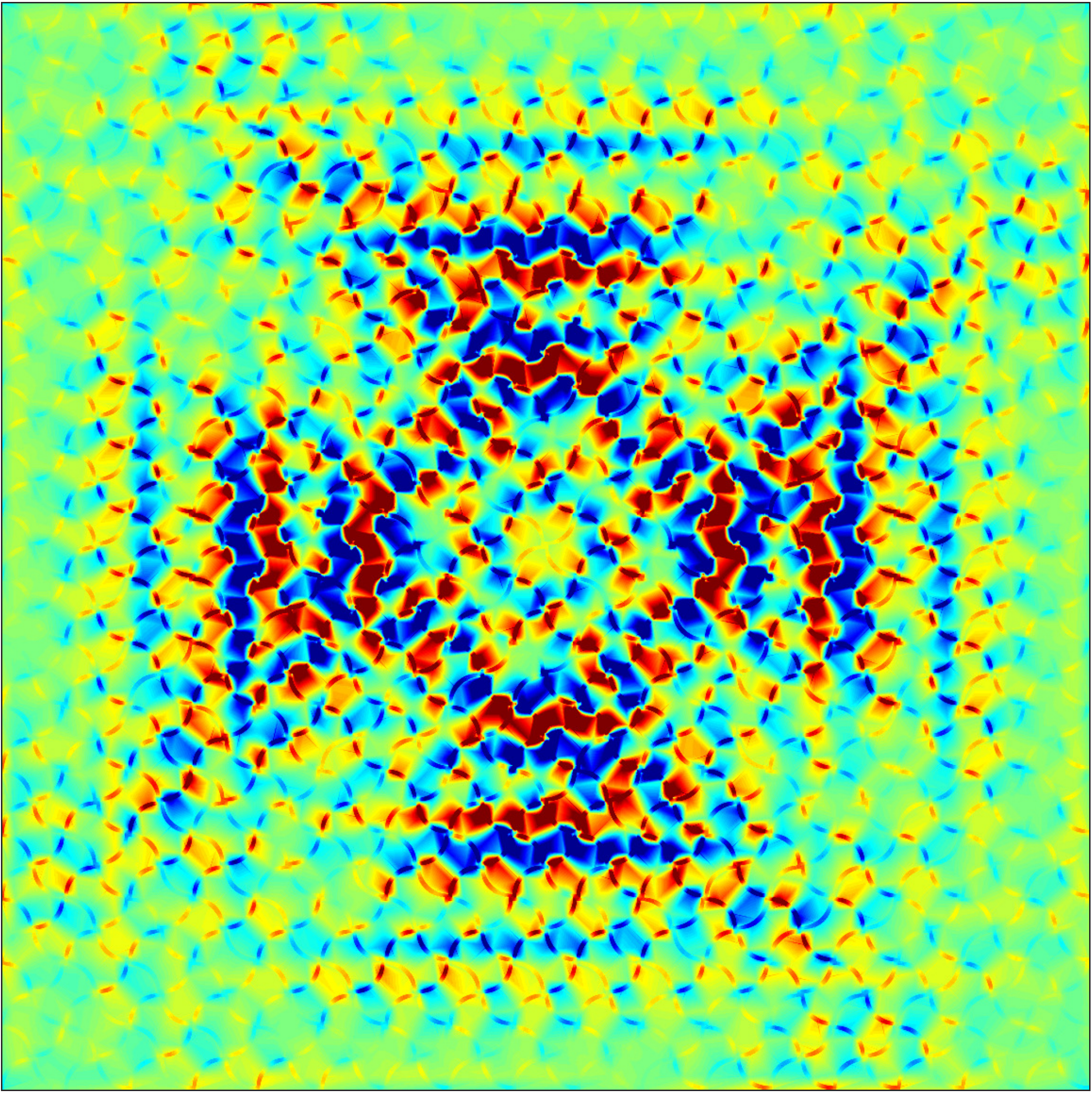}} 
\caption{Characteristics of wave propagation in the \textit{linear} curved lattice: Divergence and Curl of the wavefield (at a selected time instant) in the curved lattice for excitation frequencies at $\omega$ (a-b) and $2\omega$ (c-d), respectively.}
\label{fig52.6}
\end{figure}
\indent From the group velocity contours at $\omega$, we observe that the shear mode is highly directional, while the longitudinal mode is almost isotropic - characteristics commonly associated with in-plane wave propagation. In order to verify these spatial directivity patterns, we employ the Newmark-Beta algorithm to numerically integrate the linearized equations of motion for a finite lattice of $17\times17$ unit cells (characterized by a linearized strain-displacement relation), excited at the center with a burst applied along the horizontal and vertical directions\footnote{We assume free-free boundary conditions, and the simulation is stopped before the incident wave interacts with the boundaries}. To better elucidate the modal characteristics of the wave packet, we invoke the Helmholtz decomposition and evaluate the divergence and curl of the displacement field, which capture the longitudinal and shear mode characteristics of the wave packet, respectively \cite{Casadei2013}. Snapshots of these modal components excited at $\omega$ are plotted in figs.~\ref{fig52.6}(a)-(b), where we observe that the curl of the displacement field, which denotes shear wave behavior, is significantly activated, while the divergence component, which denotes the longitudinal wave characteristics, is almost nonexistent. This is consistent with the well-known behavior of lattice materials excited in a frequency range where longitudinal and shear modes coexist, as most of the energy associated with the excitation goes into the activation of the shear mode, which involves more compliant mechanisms of deformation. In addition, we observe that the energy mostly propagates along the diagonals from the point of excitation, consistently with the spatial characteristics determined from the group velocity contours (note that the spatial characteristics are slightly offset from the diagonal direction, which is attributed to the chirality of this lattice configuration).\newline
\indent At $2\omega$, the group velocity contours predict complementary directional patterns developing along the horizontal and vertical directions. The divergence and curl of the wavefield excited at $2\omega$ in the linearized lattice are plotted in figs.~\ref{fig52.6}(c)-(d), where the directional behavior predicted by the group velocity contours is fully recovered. Furthermore, we observe that the divergence of the wave packet is now non-negligible, which implies that this mode is characterized by both longitudinal and shear mode characteristics.\newline 
\begin{figure}[!htb]
\centering
\subfloat[Vertical displacement field excited at $\omega$]{\includegraphics[scale=0.505]{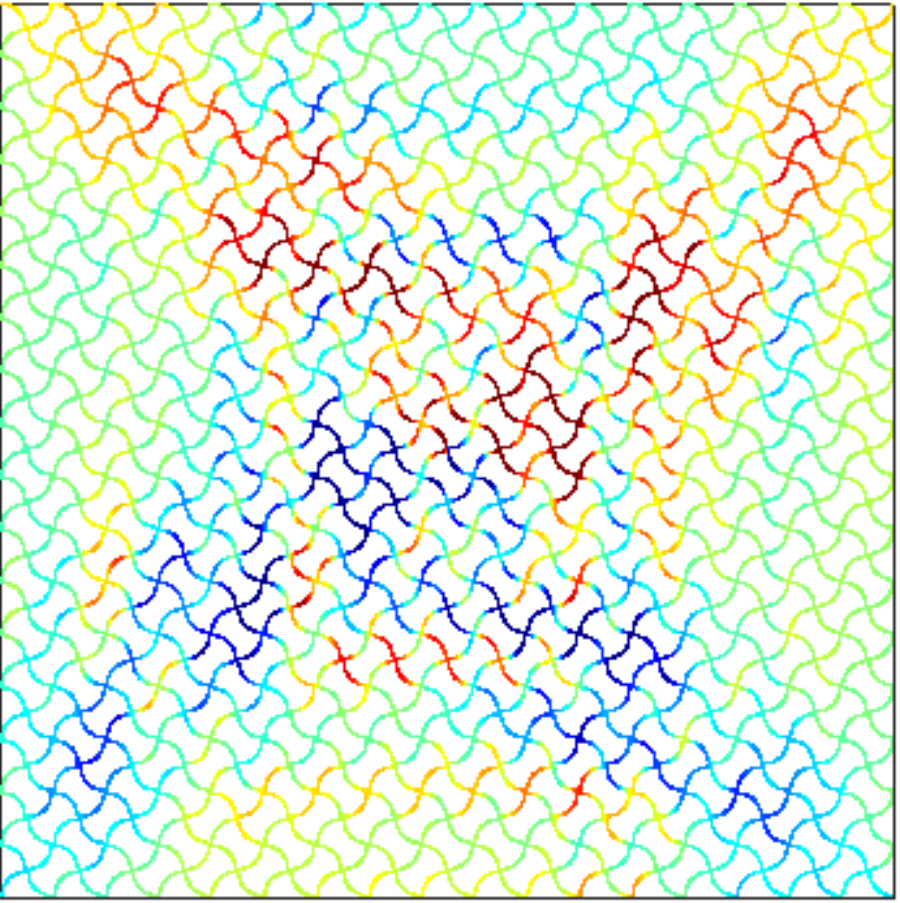}} \qquad
\subfloat[Curl of the wavefield excited at $\omega$]{\includegraphics[scale=0.2]{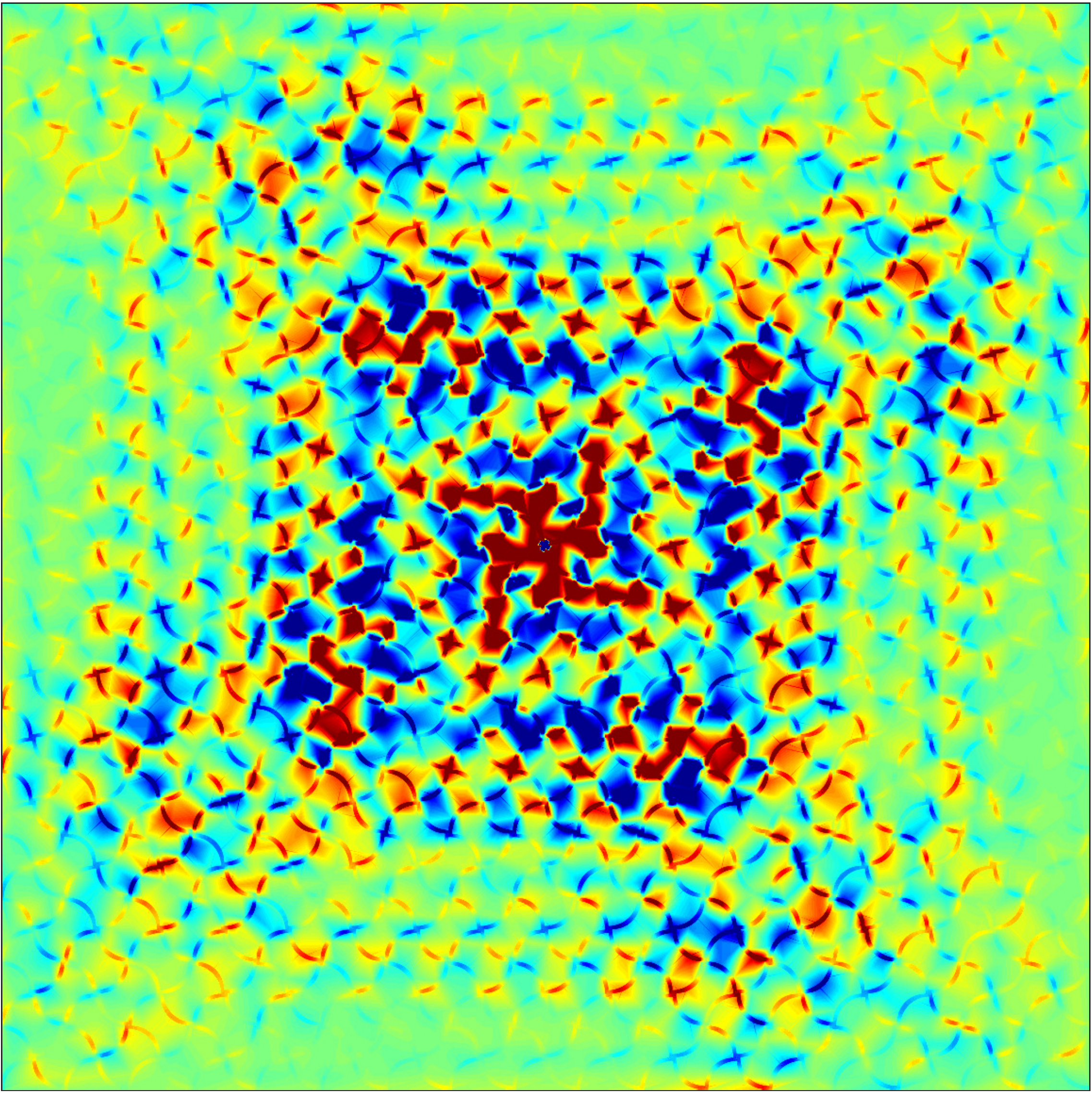}} \qquad
\subfloat[Curl of the low-pass filtered wavefield excited at $\omega$]{\includegraphics[scale=0.2]{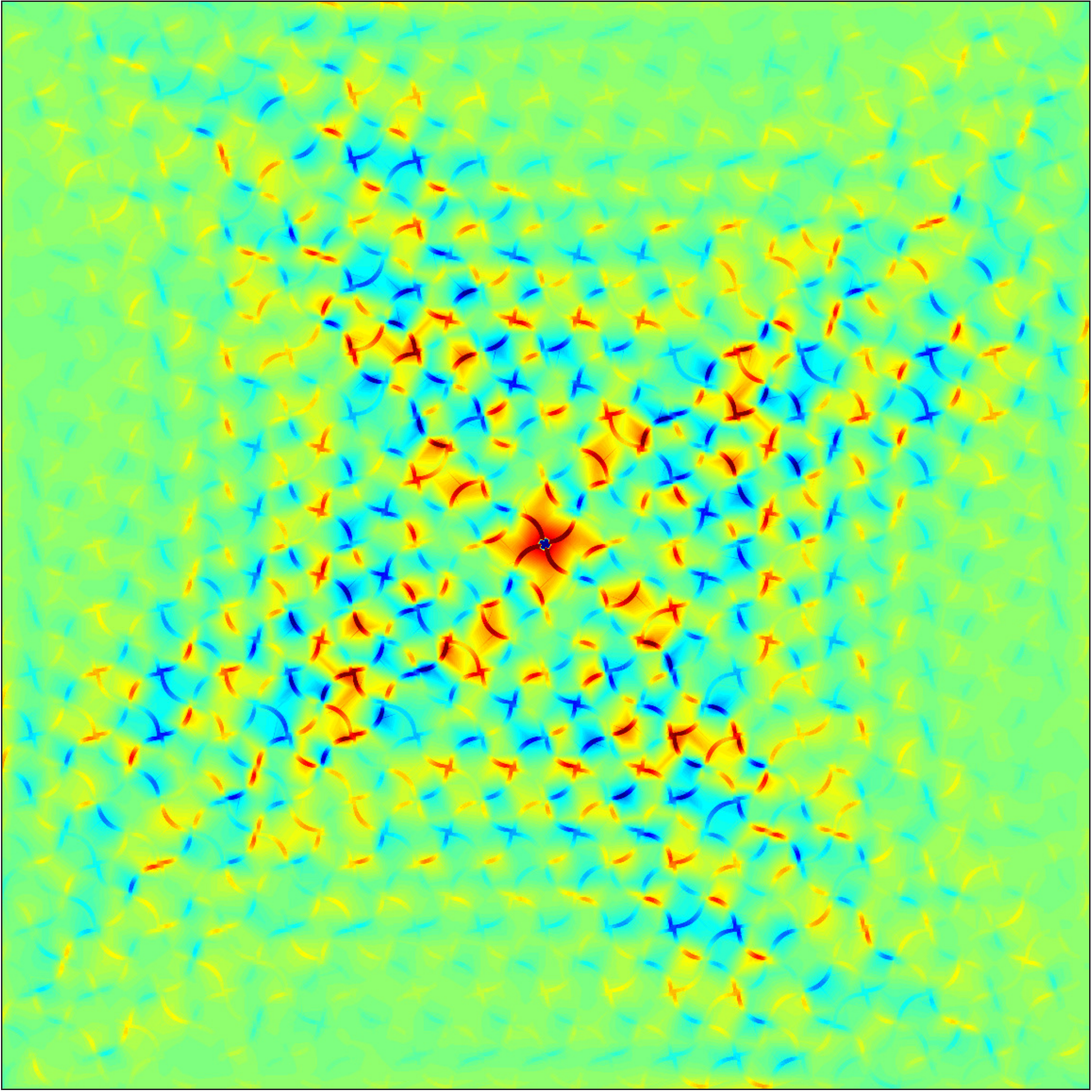}}  
\caption{Characteristics of wave propagation in the \textit{nonlinear} curved lattice : (a) Vertical displacement field for an excitation at $\omega$ (b) Curl of the wavefield excited at $\omega$ (c) Curl of the low-pass filtered wavefield, highlighting the modal characteristics of the fundamental harmonic}
\label{fig52.7}
\end{figure}
\indent Now that the linear behavior of the lattice is completely mapped out for the two frequencies, we consider the same lattice characterized by the complete nonlinear strain-displacement relationship, and we repeat the full-scale simulation (with burst excitation at $\omega$ and same boundary conditions) using an energy-momentum conserving algorithm \cite{Simo1992}. A snapshot of the (vertical) nodal displacement field is plotted in fig.~\ref{fig52.7}(a), where we observe that the amplitude in the top half of the structure features a positive offset, while the bottom half sees a negative offset. This offset is attributed to the long-wavelength modulation of the envelope of the wave packet ($A_0$ in eqn.~\ref{eq4.13}), which is a clear signature of the presence of quadratic nonlinearity in the system. In order to visualize the long-wavelength component, note that it is necessary to consider the actual displacement field, rather than the Helmholtz components, as the gradients involved in the divergence and curl operators act as \textit{de facto} high-pass filters and eliminate the long-wavelength component from the response ($A_0$ is constant with respect to the fast spatiotemporal variable).\newline
\indent Due to the weak nature of nonlinearity, we expect the modal characteristics of the nonlinear wave to be sufficiently well-represented by the modes of the corresponding linearized system (eqn.~\ref{eq4.13}). To check this hypothesis, we again consider the Helmholtz decomposition. We observe that the curl of the spatial wavefield, plotted in fig.~\ref{fig52.7}(b), features additional structure when compared with the corresponding linear response plotted in fig.~\ref{fig52.6}(b). The presence of these additional features is consistent with the complex nature of the nonlinear solution presented in eqn.~(\ref{eq4.13}), and proper identification and visualization of these different contributions necessitates additional post-processing of the spatiotemporal response. To this end, we apply a temporal low-pass/high-pass filter to the time-series data (using nodal data from the finite element simulation) in order to separate the fundamental harmonic and the nonlinearly-generated second harmonic. The curl of the low-pass filtered fundamental harmonic is plotted in fig.~\ref{fig52.7}(c), and is essentially identical to fig.~\ref{fig52.6}(b), thus confirming that the spatial characteristics of the fundamental harmonic are not affected significantly by nonlinearity in these regimes of propagation.\newline
\begin{figure}[!htb]
\centering
\subfloat[Divergence of the high-pass filtered wavefield excited at $\omega$]{\includegraphics[scale = 0.25]{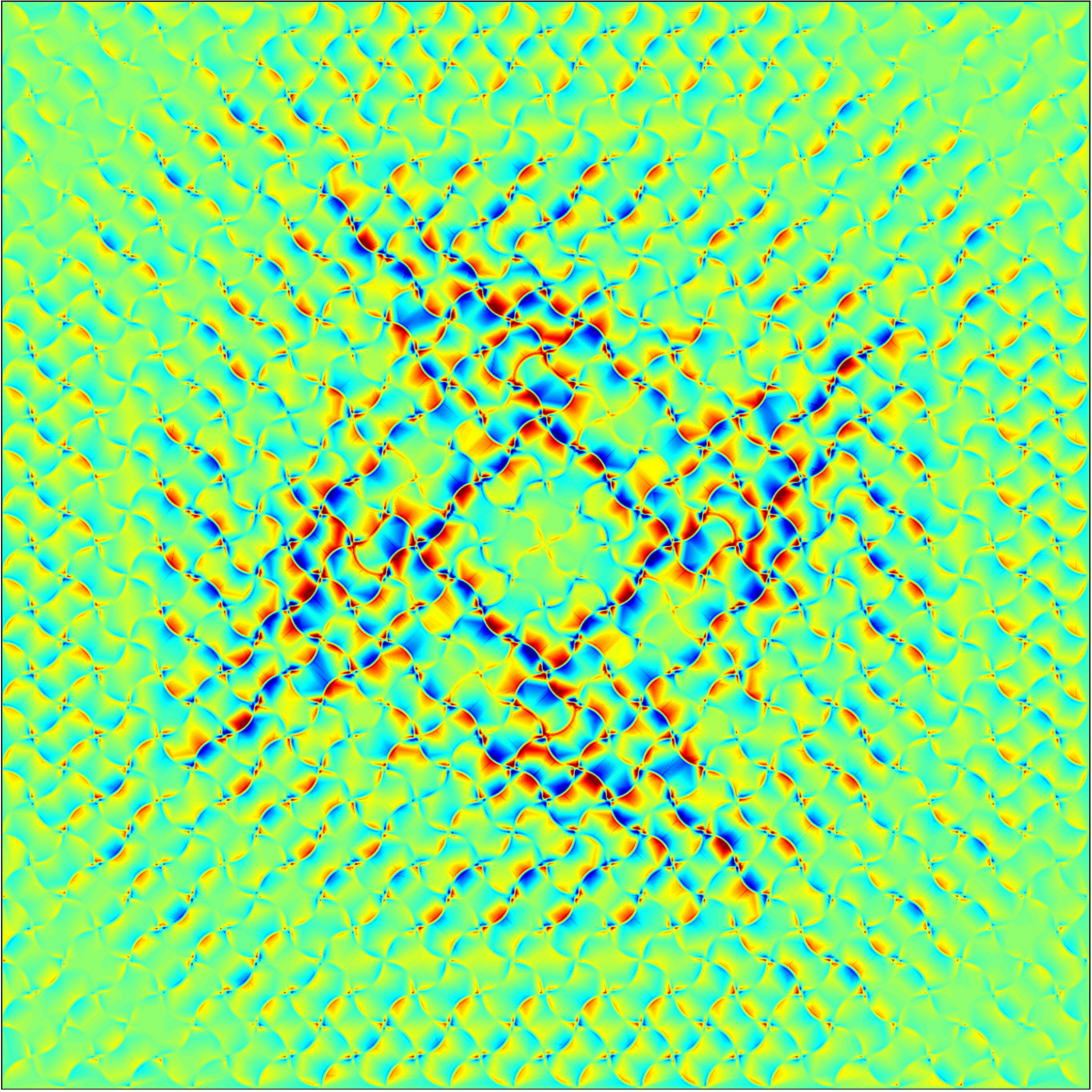}} \qquad\qquad
\subfloat[Divergence of the linear wavefield excited at $2\omega$]{\includegraphics[scale=0.25]{Curv2_L2_div}} 
\caption{Divergence of the high-pass filtered \textit{nonlinear} wavefield, highlighting the modal characteristics of the nonlinearly-generated second harmonic (a) compared with the divergence of the wavefield obtained in the \textit{linear} lattice excited at $2\omega$ (b).}
\label{fig52.8}
\end{figure}
\indent As we know from fig.~\ref{fig52.6}(c) that the linear wave at $2\omega$ also activates longitudinal mode characteristics, we monitor the divergence of the high-pass filtered data, a snapshot of which is plotted in fig.~\ref{fig52.8}(a). As predicted by eqn.~(\ref{eq4.13}), we observe a wave profile with modal characteristics that are remarkably consistent with the predictions obtained from the linearized dispersion relation (and the linear simulations) at $2\omega$ (reported again for convenience in fig.~\ref{fig52.8}(b)); the additional spatial components in fig.~\ref{fig52.8}(a) can be attributed to the forced component corresponding to the second harmonic in eqn.~(\ref{eq4.13}). Therefore, by merely controlling the amplitude of the excitation, we can establish complementary directional characteristics with respect to those of the fundamental harmonic. In addition, we also note that longitudinal characteristics (in general, deformation characteristics typically associated with higher frequency modes) are now excited in the nonlinear lattice at frequencies where the corresponding linear lattice would only display shear wave characteristics.\newline
\indent In summary, the manifestations of nonlinearity described by eqn.~(\ref{eq4.13}), namely amplitude modulation and harmonic generation, are recovered from the numerical simulation of complex lattices characterized by an equivalent quadratic nonlinearity. These observations demonstrate the generality of the solutions obtained in section \ref{sec2} and serve as a proof of concept for the design of metamaterial architectures that are capable of adaptively augmenting their directivity landscape and activating functionalities that are seldom accessible at lower frequencies of excitation. In the next example, we will extend this proof of concept to a lattice configuration displaying both geometric and material nonlinearity.
\subsection{Soft Cellular Solid}
\begin{figure}[!htb]
\centering
\includegraphics[scale=0.25]{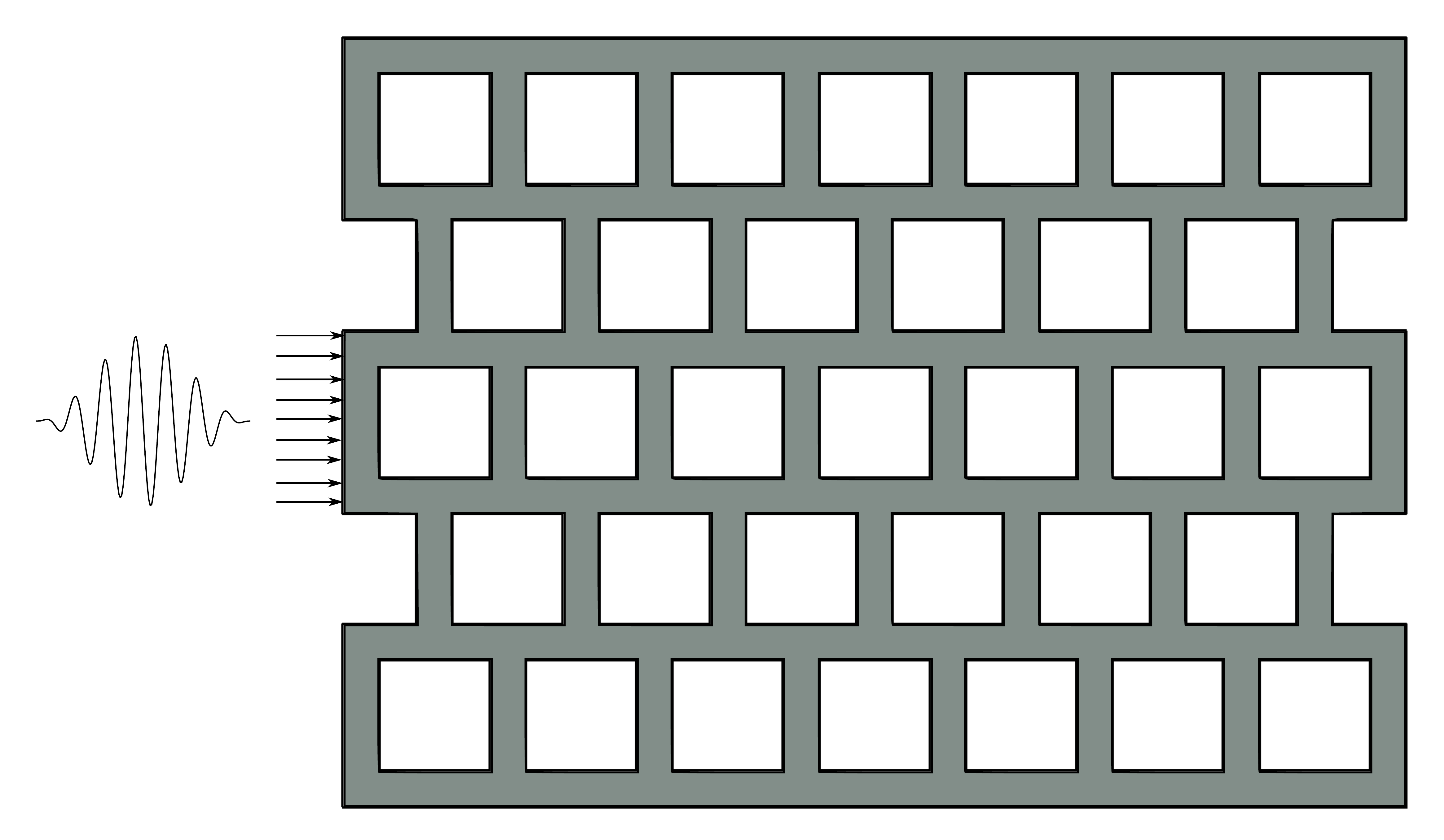} 
\caption{Schematic of Soft, brick wall based metamaterial excited harmonically at the center of the left edge.}
\label{fig62.1}
\end{figure}

The nonlinear lattice under consideration consists of a square unit cell tessellated in a brick wall (staggered) pattern as shown in fig.~\ref{fig62.1}. In addition to geometric nonlinearity, we assume that the stress-strain relationship is also nonlinear (material nonlinearity), and can be expressed using a Neo-Hookean constitutive model as 
\begin{equation}
\mathcal{W} = \frac{\lambda}{8}\,(log \mathcal{I}_3)^2 + \frac{\mu}{2}\,(\mathcal{I}_1-3-log (\mathcal{I}_3)),
\end{equation}
where $\lambda$ and $\mu$ are the associated Lam\'e parameters (chosen to be the same as those used for the curved lattice). While a number of constitutive relations are available in literature to capture the physical behavior of soft materials, we choose the Neo-Hookean model as a generic test bed to demonstrate the effect of material nonlinearity on wave propagation. Furthermore, we assume a state of plane strain in the lattice in order to reduce the constitutive equations to two dimensions.\newline
\begin{figure}[!htb]
\centering
\subfloat[Band diagram]{\includegraphics[scale=0.7]{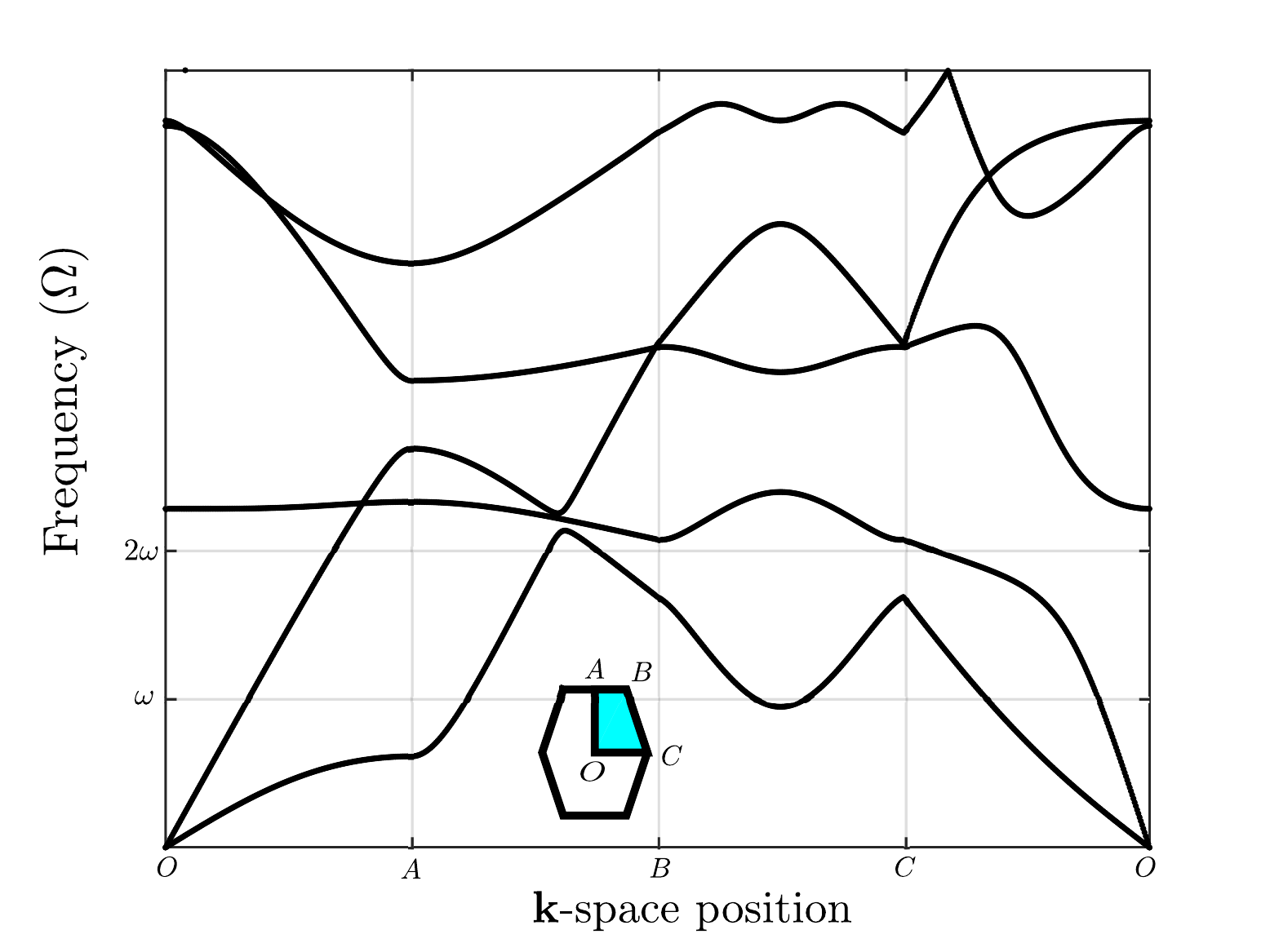}} \\
\subfloat[Group velocity contours at $\omega$]{\includegraphics[scale=0.45]{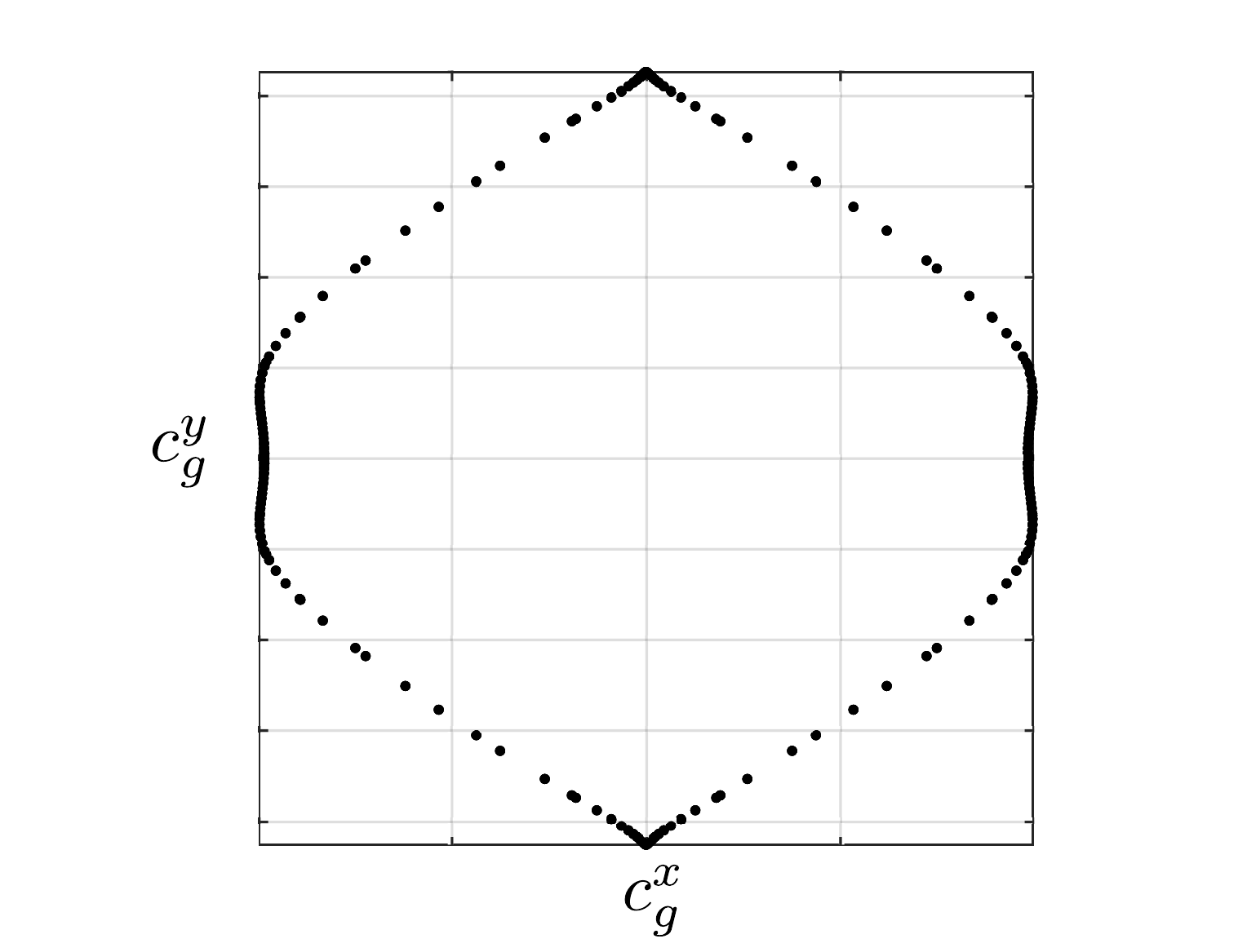}} \qquad \qquad
\subfloat[Group velocity contours at $2\omega$]{\includegraphics[scale=0.45]{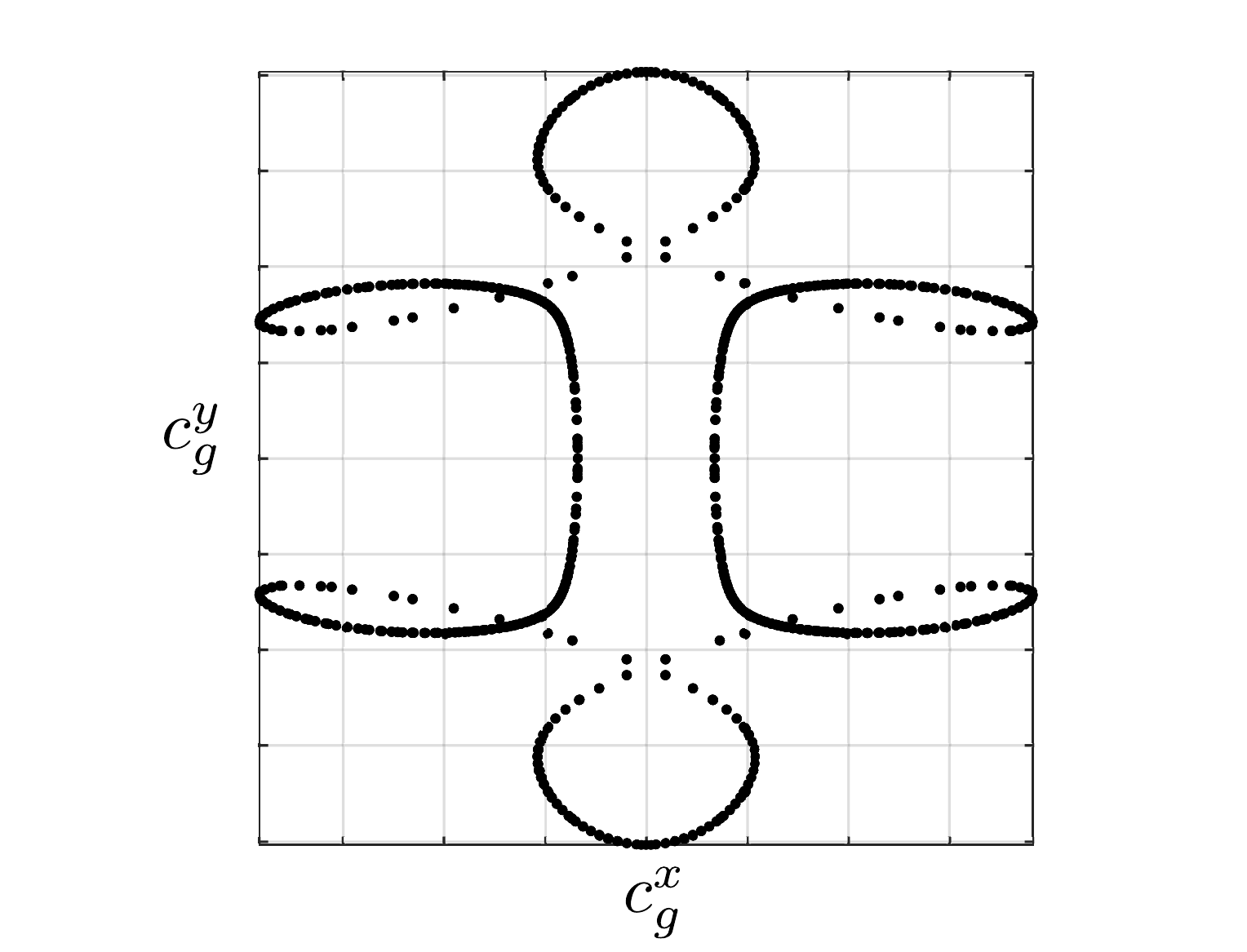}}
\caption{Dispersion characteristics of the linearized lattice: (a) Band structure of the lattice evaluated using $2$D plane strain elasticity. (b)-(c) Group velocity contours for the longitudinal mode evaluated at $\omega$ and $2\omega$, respectively.}
\label{fig62.2}
\end{figure}
\indent For small displacements, the Neo-Hookean model can be linearized to the St. Venant-Kirchoff model, which is then employed to determine the band structure and group velocity characteristics of linear wave propagation\footnote{Alternately, a small-on-large technique can also be employed to numerically determine the tangent stiffness at any static finite deformation, which can then be used for linear eigen analysis \cite{Bertoldi2008}. In this case, both techniques yield the same result.} (shown in fig.~\ref{fig62.2}). For our analysis, we choose two frequencies $(\omega,2\omega)$ which live on the same dispersion branches, but excite the corresponding modes in distinct wavenumber regimes. Since we plan to excite the lattice in a way that primarily engages the longitudinal modes, we focus our attention on the group velocity contours of the longitudinal mode, which are plotted in figs.~\ref{fig62.2}(b)-(c) for excitations centered at $\omega$  and $2\omega$, respectively. \newline
\indent At $\omega$, the group velocity contours predict a plane wave front that propagates along the horizontal direction, as well as some lateral focussing along the vertical direction, while a wide spectrum of diagonal directions are essentially de-energized. The horizontal plane wave features are attributed to the fact that the wavelength of the excitation is much larger than the length of the unit cell, thereby resulting in a response that is immune to the presence of unit cells in the staggered direction and predominantly travels along the available horizontal waveguides. At $2\omega$, the wavelength becomes shorter and the lattice begins to display dispersive behavior even along the horizontal direction, marked by the appearance of new caustics in the velocity contours, as shown in fig.~\ref{fig62.2}(c). These caustics indicate the introduction of two new pathways for wave propagation in the lattice, that are not exhibited by the same longitudinal modes at lower frequencies.\newline
\begin{figure}[!htb]
\centering
\subfloat[Radial component of wavefield excited at $\omega$]{\includegraphics[scale=0.25]{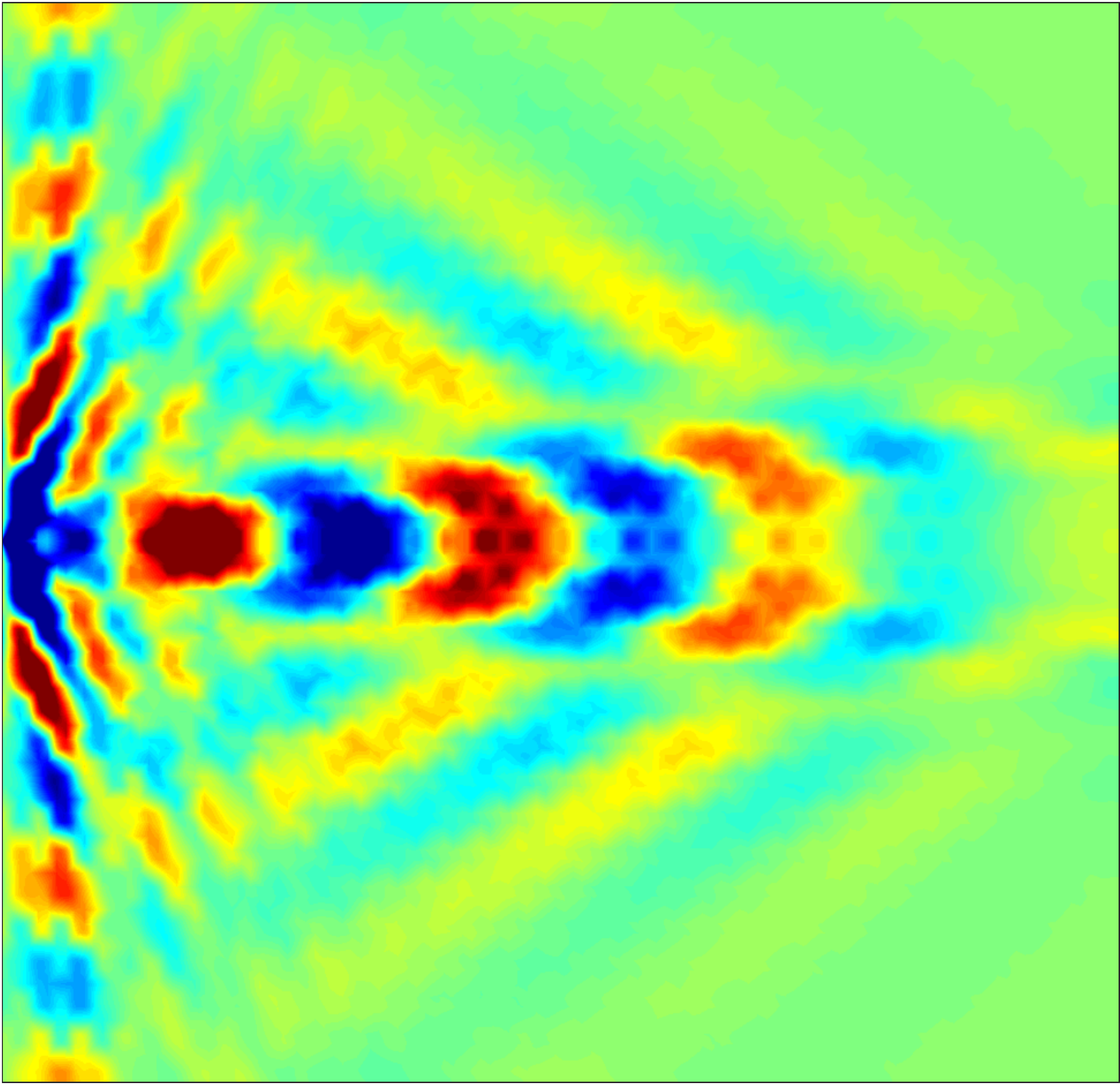}} \qquad\qquad 
\subfloat[Spectral contours of the wavefield in (a)]{\includegraphics[scale=0.23]{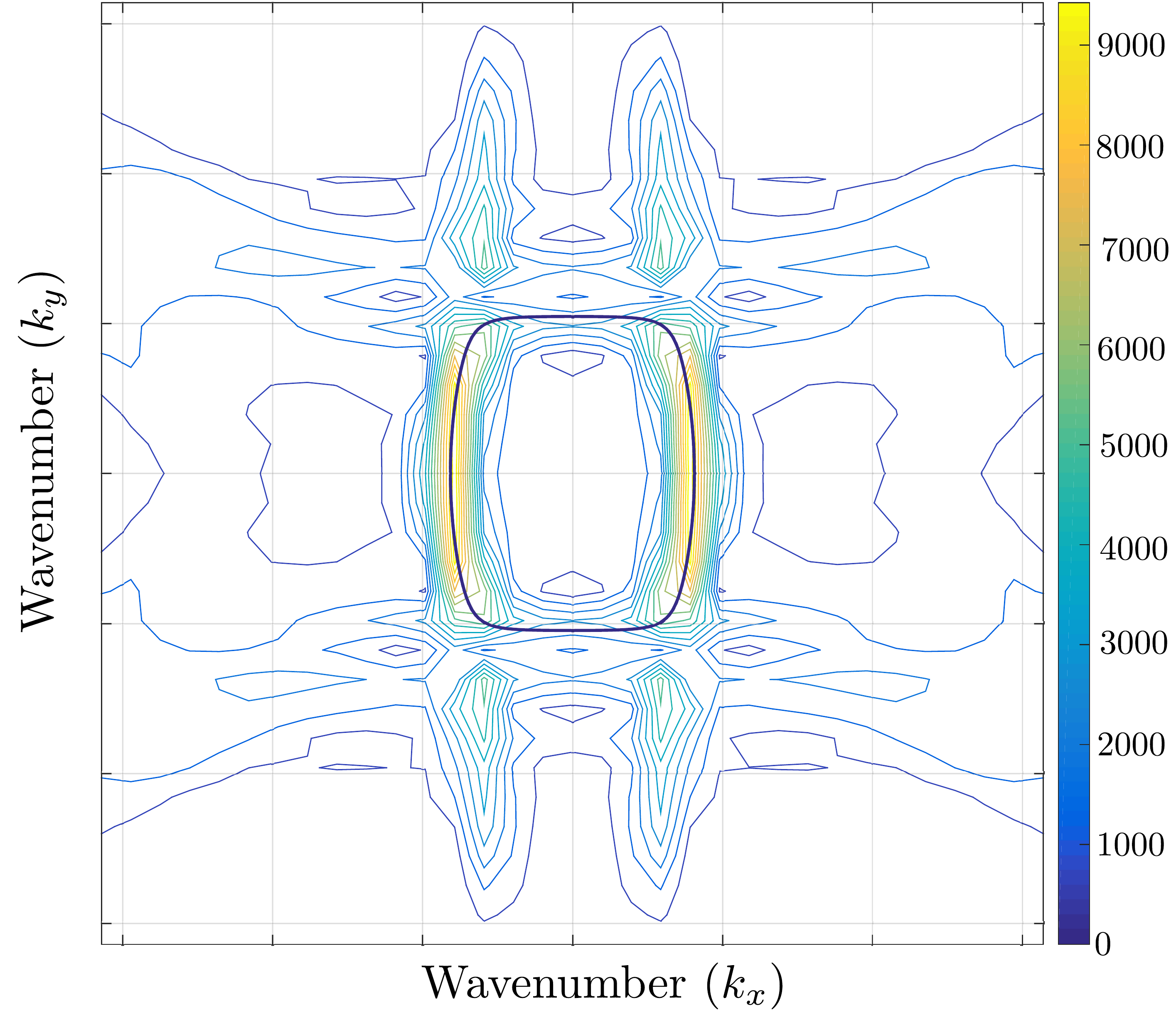}} \\ \vspace{0.05in}
\subfloat[Radial component of wavefield at $2\omega$]{\includegraphics[scale=0.25]{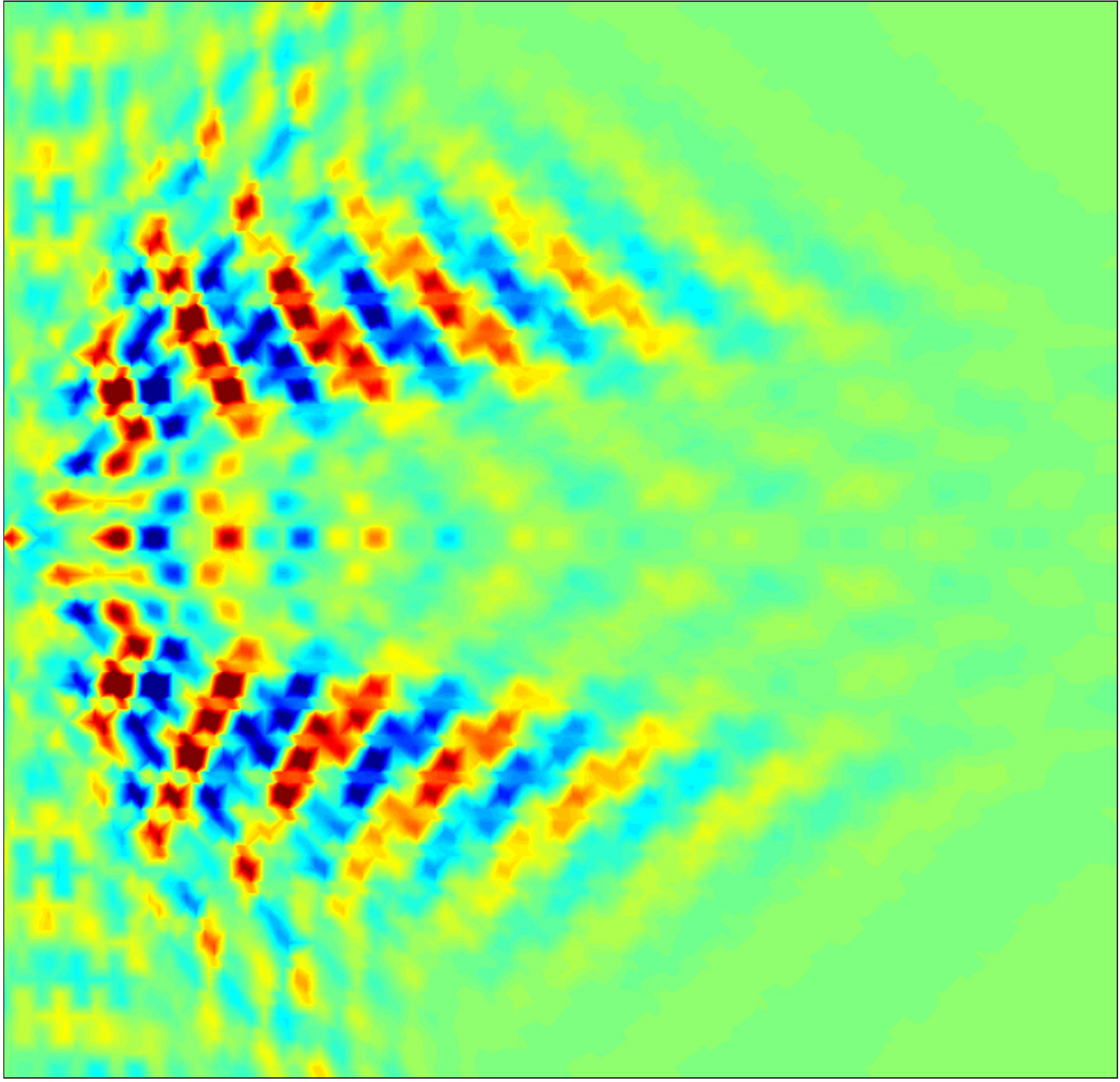}} \qquad\qquad
\subfloat[Spectral contours of the wavefield in (c)]{\includegraphics[scale = 0.23]{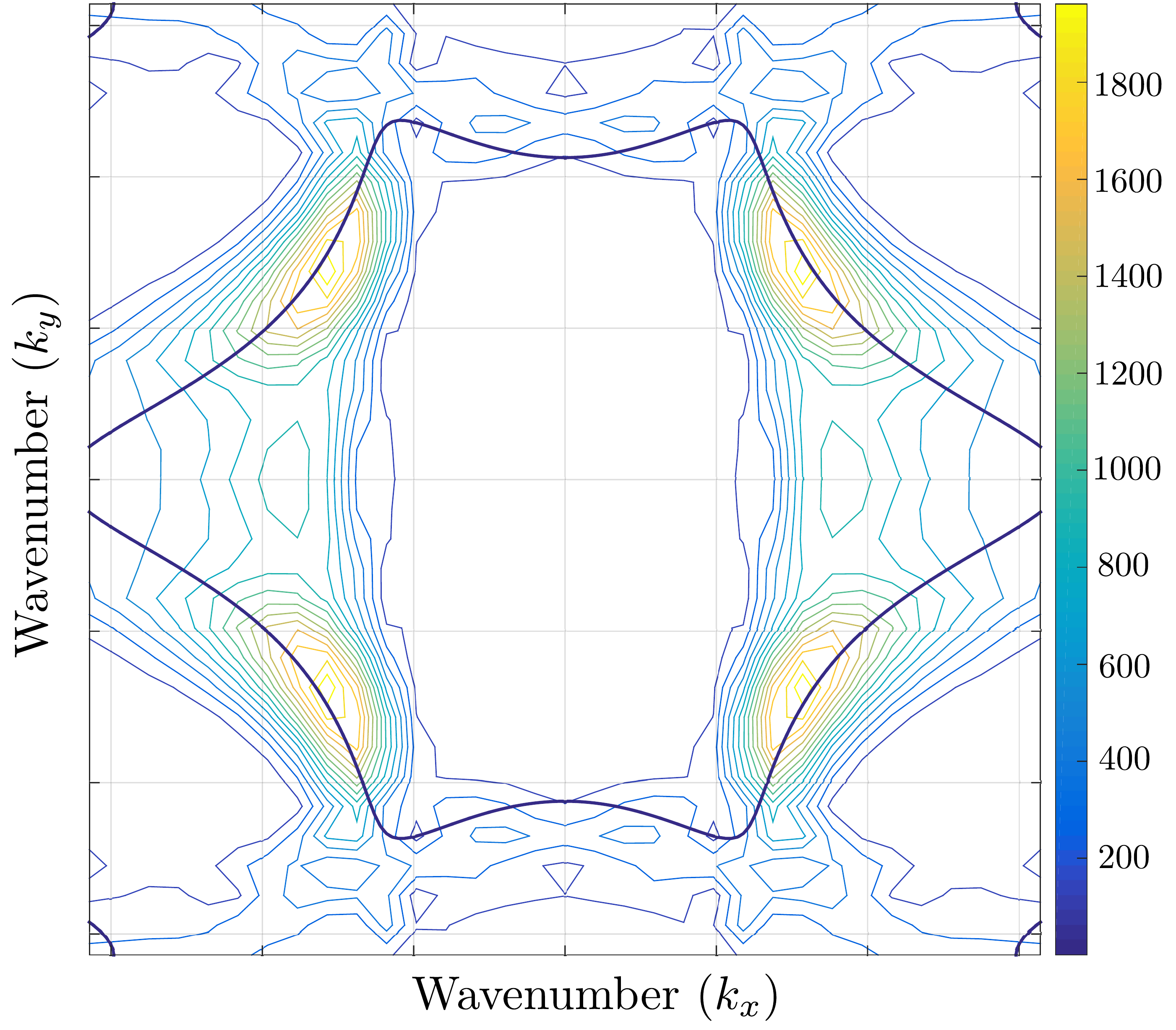}} \\
\caption{Characteristics of wave propagation in the \textit{linearized} brick metamaterial: Radial component of the spatial wavefield (and the corresponding 2-D DFT) for excitation frequencies $\omega$ (a-b) and $2\omega$ (c-d) respectively.}
\label{fig62.3}
\end{figure}
\indent We then proceed to simulate wave propagation in a finite lattice, excited horizontally at the center of the left edge as shown in fig.~\ref{fig62.1}. To verify the spatial directivity patterns obtained from the unit cell analysis, we first perform full-scale simulations of the linearized lattice. We plot the wavefield in polar coordinates (radial and tangential components)\footnote{The origin of this system lies at the center of the left edge of the finite lattice.}, and closely inspect the radial component, which captures the longitudinal characteristics of the wave. The radial component for the excitation centered at $\omega$ is plotted in fig.~\ref{fig62.3}(a), where we observe that the energy predominantly propagates along the horizontal direction, consistently with the predictions from the group velocity contours (a minor contribution also propagates along the edges of the lattice in the vertical direction). In fig.~\ref{fig62.3}(b), we plot the $2$D DFT of the spatial wavefield superimposed to the isofrequency contours of the longitudinal mode, which further confirms the agreement between unit cell analysis and full-scale simulations and showcases the quasi-plane wave nature of the propagating wavefield.\newline
\begin{figure}[!htb]
\centering
\subfloat[Radial component of wavefield excited at $\omega$]{\includegraphics[scale=0.25]{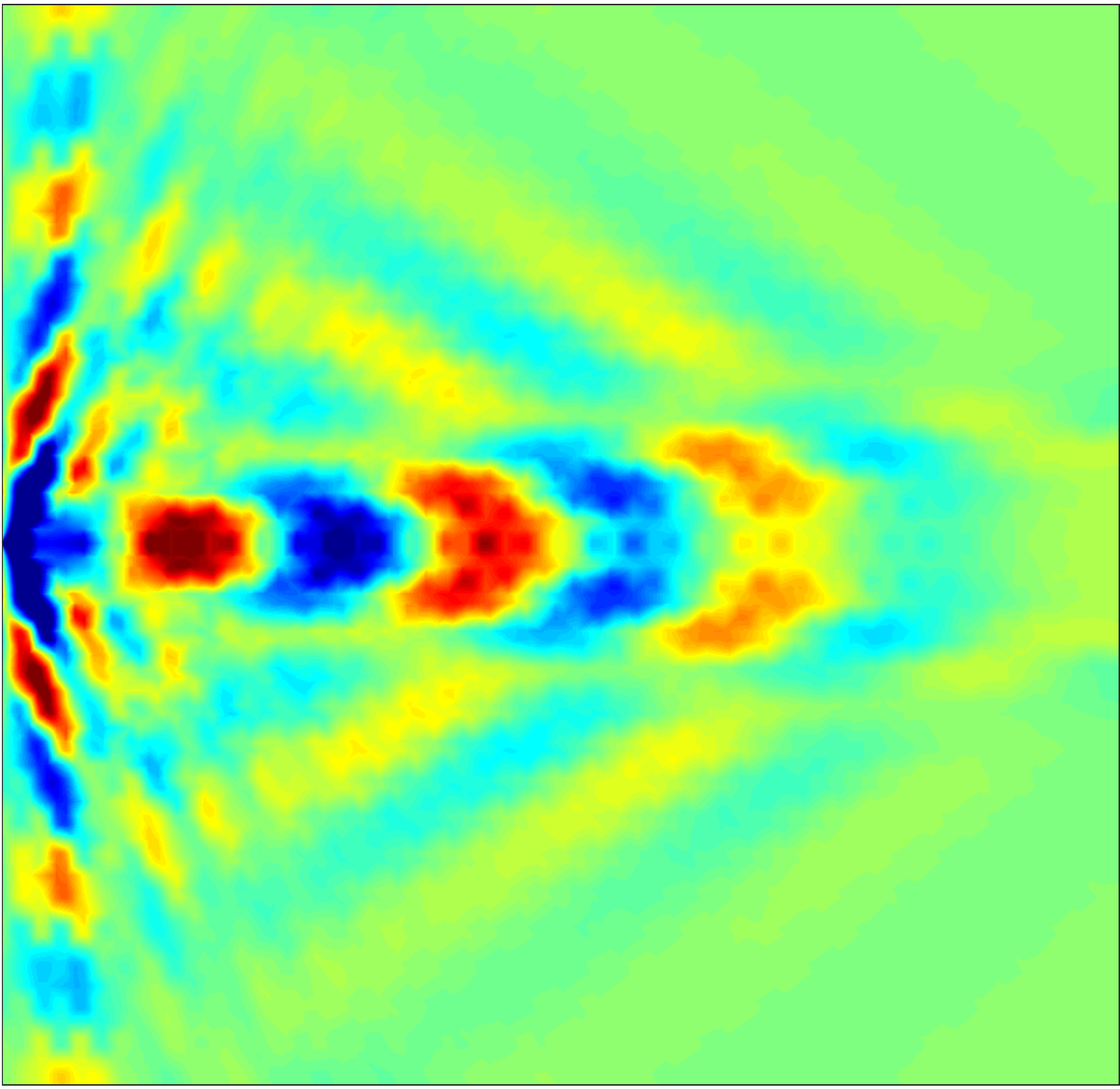}} \qquad\qquad
\subfloat[Spectral contours of the wavefield in (a)]{\includegraphics[scale=0.23]{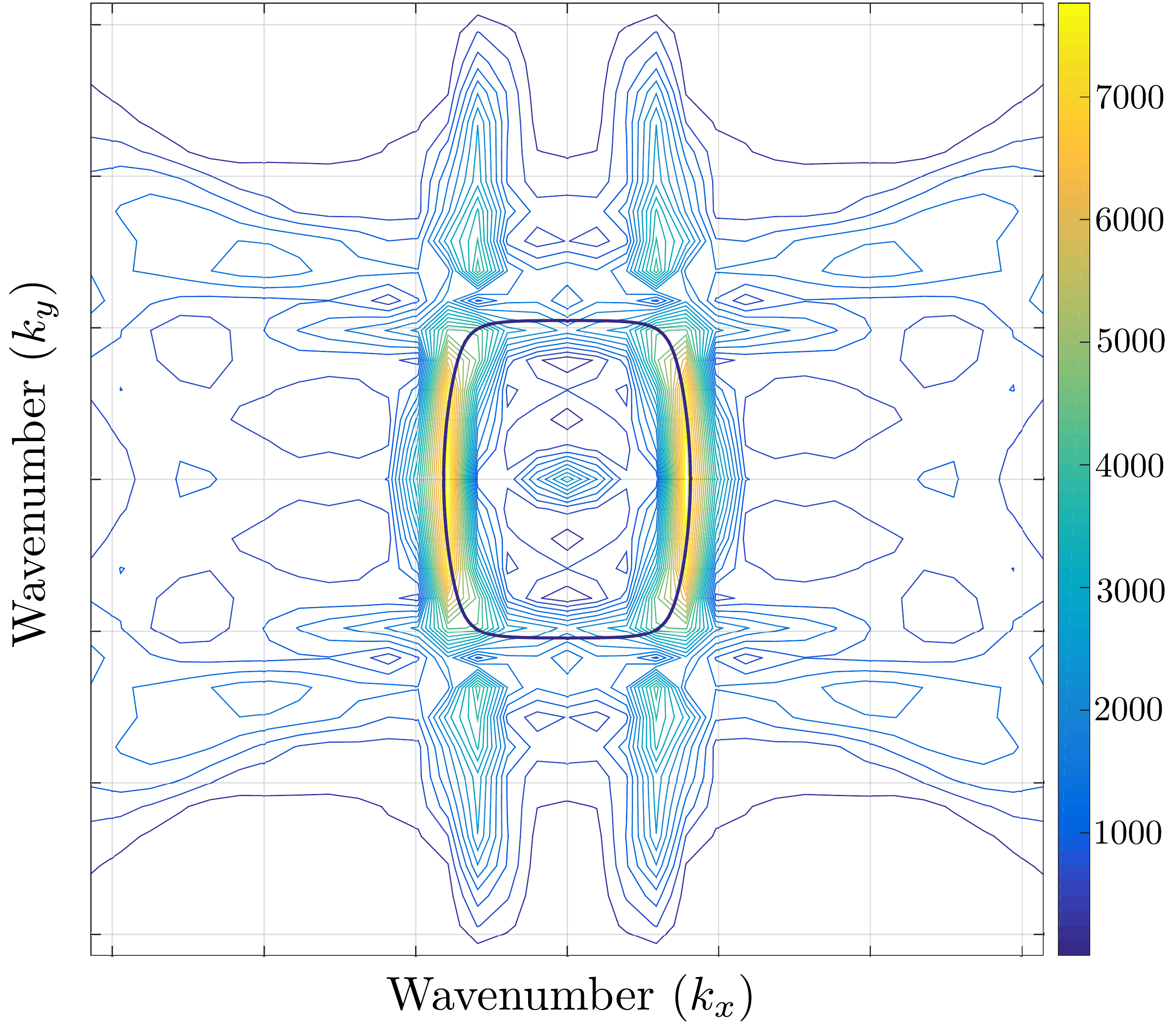}} \\ \vspace{0.05in}
\subfloat[Radial component of the high-pass filtered wavefield excited at $\omega$]{\includegraphics[scale=0.25]{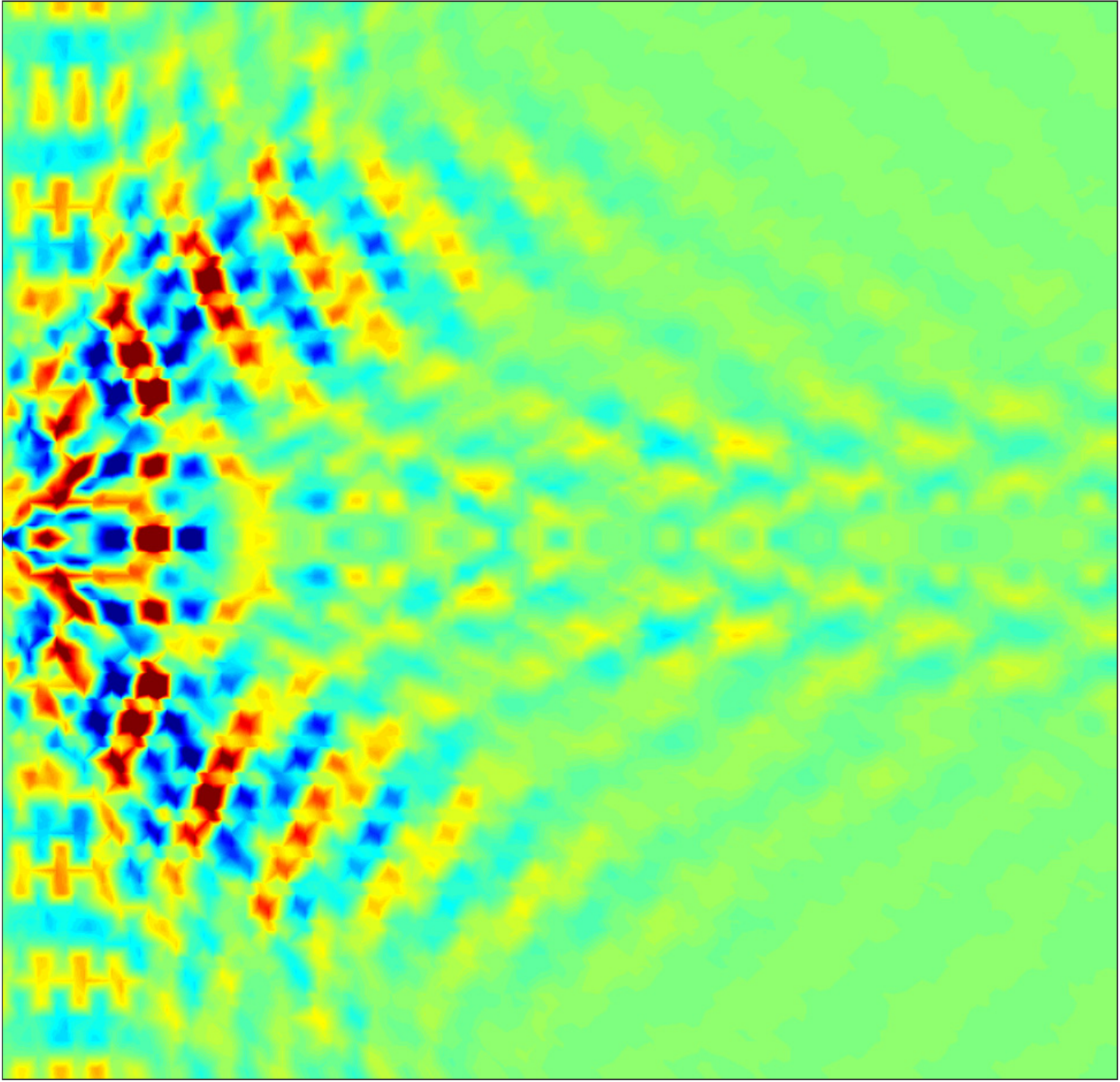}} \qquad \qquad
\subfloat[Spectral contours of the wavefield in (c)]{\includegraphics[scale = 0.23]{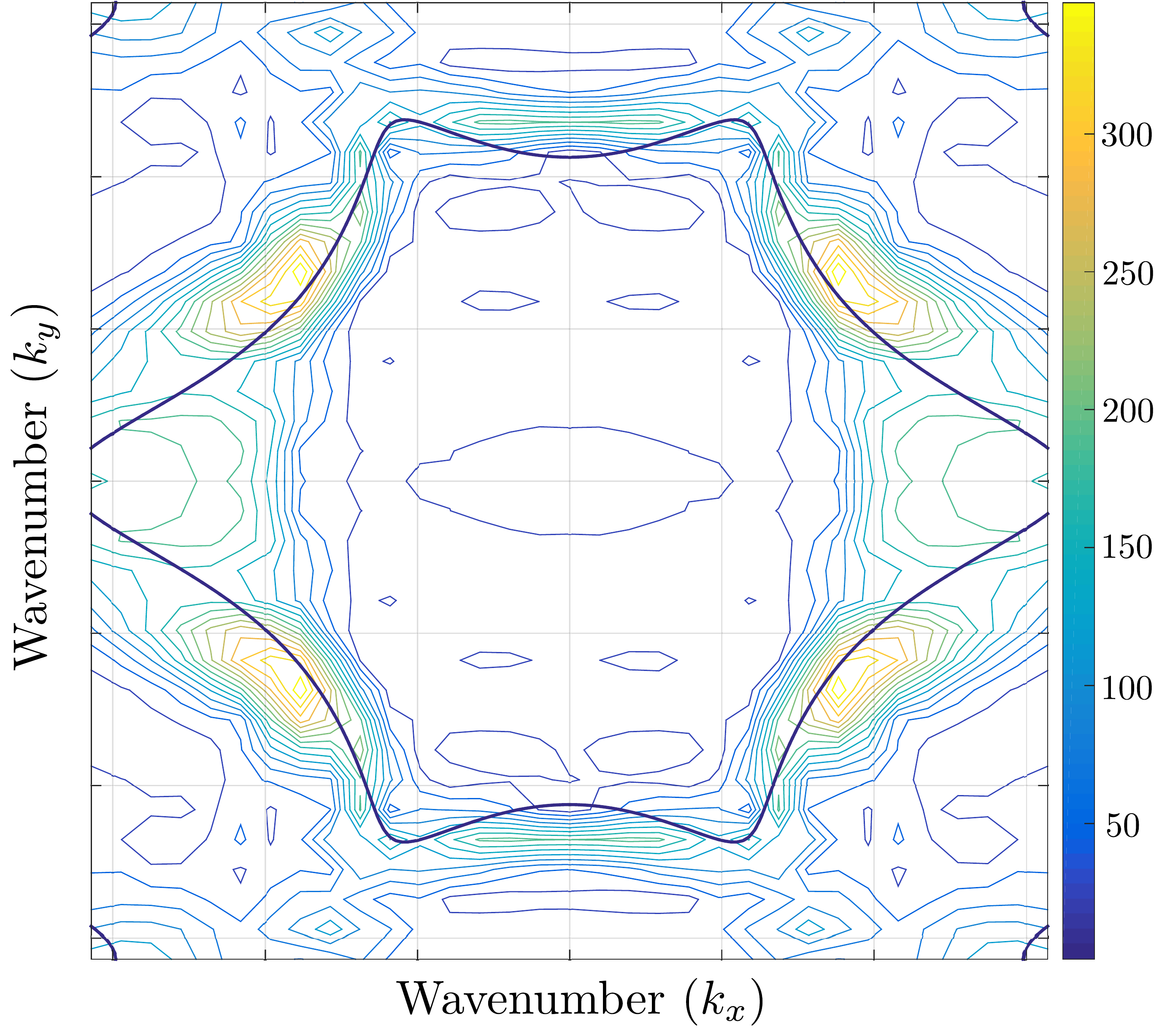}} \\
\caption{Characteristics of wave propagation in the \textit{nonlinear} brick metamaterial: (a) Radial component of the wavefield (b) Spectral contours of the radial component revealing the presence of a long-wavelength envelope modulation. (c-d) Radial component of the high-pass filtered wavefield and its corresponding spectral representation showing complementary diagonal energy re-direction.}
\label{fig62.4}
\end{figure}
\indent For the excitation at $2\omega$, the spatial and spectral components of the wavefield plotted in fig.~\ref{fig62.3}(c)-(d) also match the group velocity contours accurately. Thus, from the linear analysis of the two frequencies, we conclude that the lower frequency corresponds to a quasi plane-wave propagation, aligned with the horizontal waveguides and the direction of excitation, while the higher frequency gives rise to a directional pattern that splits the wavefield into two distinct diagonal fronts, away from the horizontal waveguide direction.\newline
\indent Finally, we consider the same finite lattice, now exhibiting material and geometric nonlinearity, and repeat the full-scale simulations for an excitation centered at $\omega$. The radial component of the wavefield plotted in fig.~\ref{fig62.4}(a) bears a striking similarity to the linear wavefield plotted in fig.~\ref{fig62.3}(a). However, when we inspect the spectral characteristics of the response, we observe an additional long-wavelength component (centered around $k_x=k_y=0$), which can be attributed once again to the manifestation of the equivalent quadratic nonlinearity (term $A_0$ in eqn.~\ref{eq4.13}). In order to visualize the nonlinearly-generated harmonics, we apply a high-pass filter to the spatiotemporal displacement history, and the radial component of the wavefield is plotted in fig.~\ref{fig62.4}(c), where we observe that the energy associated with the second harmonic indeed propagates away from the primary waveguide direction. In addition, the spectral response of the filtered second harmonic plotted in fig.~\ref{fig62.4}(d) matches the isofrequency contours at $2\omega$, which confirms that the homogeneous characteristics of the nonlinearly-generated harmonics are indeed described accurately by those of the corresponding linear system. Therefore, the nonlinear lattice gives rise to the introduction of directional wave characteristics at frequencies where the corresponding linear lattice would only exhibit horizontal plane wave features. This system can therefore be seen as the continuum 2D analog of the adaptive longitudinal-flexural energy converter that we previously proposed in \cite{GG15}, in which the energy of high-amplitude excitations was rerouted laterally away from the main waveguide direction. 
\section{Conclusion}
In this work, we have shown that the interplay of nonlinearity and dispersion involved in the generation of higher harmonics is responsible for unique characteristics seldom observed in nondispersive systems. Specifically, the presence of modal complexity in the structure leads to the generation of higher harmonics with a variety of spectro-spatial characteristics. In particular, one component of the higher harmonic conforms to the dispersion characteristics of the linearized system. As a result, the characteristics of this harmonic can be determined from a simple eigen analysis of the corresponding linear structure. Therefore, we conclude that, by simply analyzing the linear wave characteristics of a periodic structure, we can predict (or even design) the modal functionalities that can be activated through the presence of nonlinear mechanisms, regardless of the source of nonlinearity in the system (geometric, material, and contact nonlinearity). We have demonstrated the generality of this design paradigm using the illustrative example of nonlinear lattices, where the generation of harmonics leads to the activation of complementary functionalities normally associated with high frequency modes even in response to excitations applied in the low frequency regime. Furthermore, this phenomenon often results in an enhanced wave propagation landscape in which optical cell deformation mechanisms are activated and the augmented directivity forces some of the energy to propagate along directions that are not energized in the linear regime. Finally, the ability of the nonlinear lattices to support finite deformations can also be utilized to tune the characteristics of linear wave propagation, thereby giving rise to virtually endless opportunities to engineer materials with desired tunable and switchable functionalities.
\section*{Acknowledgements}
The authors acknowledge the support of the National Science Foundation (CAREER Award CMMI-1452488).
\appendix
\section{Multiple Scales formulation for a diatomic spring-mass chain}
\label{appA}
The solution to eqn.~(\ref{eq4.1}) is written using a perturbative expansion as 
\begin{equation} 
\label{eq4.2A}
\bfu_n = \begin{Bmatrix}
u_{n}(t) \\
v_{n}(t)
\end{Bmatrix} = \begin{Bmatrix}
\sum_{i=0}^{\infty} \varepsilon^i u^{i}(\theta_n,\xi_n,\tau)\\
\sum_{i=0}^{\infty} \varepsilon^i v^{i}(\theta_n,\xi_n,\tau)
\end{Bmatrix}.
\end{equation}
Using this representation, the total derivative with respect to $t$ can be expressed using the assumption of independence of the multiple scales variables as 
\begin{align}
\frac{d()}{dt} &= \frac{\partial ()}{\partial \theta_n}\frac{d \theta_n}{dt} + \frac{\partial ()}{\partial \xi_n}\frac{d \xi_n}{dt} + \frac{\partial ()}{\partial \tau}\frac{d \tau}{dt} \nonumber \\ 
&= -\omega \frac{\partial ()}{\partial \theta_n} - \varepsilon c_g \frac{\partial ()}{\partial \xi_n} + \varepsilon^2 \frac{\partial ()}{\partial \tau} \label{eqApA2}. 
\end{align}
The multiple scales representation also results in the introduction of new variables $\theta_{n\pm1}$ and $\xi_{n\pm1}$ in the expansion, which need to be eliminated. In this regard, the slowly-varying spatiotemporal variable $\xi$ represents a long-wavelength component, which can be exploited to perform a truncated Taylor series expansion of the displacements in terms of $\xi_{n}$ as 
\begin{align}
u_{n\pm1}^{i}(\theta_{n\pm1},\xi_{n\pm1},\tau) &= u_{n\pm1}^{i}(\theta_{n\pm1},\xi_{n},\tau) \pm \varepsilon \frac{du_{n\pm1}^{i}}{d\xi_n}(\theta_{n\pm1},\xi_{n},\tau) + \frac{1}{2}\varepsilon^2 \frac{d^2u_{n\pm1}^{i}}{d\xi_n^2}(\theta_{n\pm1},\xi_{n},\tau)  \nonumber \\
v_{n\pm1}^{i}(\theta_{n\pm1},\xi_{n\pm1},\tau) &= v_{n\pm1}^{i}(\theta_{n\pm1},\xi_{n},\tau) \pm \varepsilon \frac{dv_{n\pm1}^{i}}{d\xi_n}(\theta_{n\pm1},\xi_{n},\tau) + \frac{1}{2}\varepsilon^2 \frac{d^2v_{n\pm1}^{i}}{d\xi_n^2}(\theta_{n\pm1},\xi_{n},\tau) \label{eqApA3}. 
\end{align}
Substituting eqns.~\ref{eq4.2A}-\ref{eqApA3} in eqn.~(\ref{eq4.1}), and separating the terms by the order of perturbation expansion $\varepsilon^i$, we obtain eqn.~(\ref{eq4}). \newline 
To determine the homogeneous solution to the equation at each order of expansion, Bloch-Floquet conditions are then used to simplify the dependence of the neighboring displacements $(u_{n\pm1}, v_{n\pm1})$, in terms of $u_n\, v_n$ as,
\begin{align}
 u_{n\pm1}(\theta_{n\pm1},\xi_{n},\tau) &= e^{\pm ik}u_n(\theta_{n},\xi_{n},\tau) \\
 v_{n\pm1}(\theta_{n\pm1},\xi_{n},\tau) &= e^{\pm ik}v_n(\theta_{n},\xi_{n},\tau).
\end{align}

 \section*{References}


\bibliographystyle{elsarticle-harv}
\bibliography{GaneshGonella_MixingLattice_JMPS}

\end{document}